\documentclass[11pt]{article}%
\usepackage{amssymb}
\usepackage[singlespacing]{setspace}
\usepackage{amsmath}
\usepackage{amsfonts}
\usepackage{graphicx}
\usepackage{cite}%
\allowdisplaybreaks
\begin{document}

\author{\vspace{0.16in}Hartmut Wachter\thanks{e-mail:
Hartmut.Wachter@physik.uni-muenchen.de}\\Max-Planck-Institute\\for Mathematics in the Sciences\\Inselstr. 22, D-04103 Leipzig\\\hspace{0.4in}\\Arnold-Sommerfeld-Center\\Ludwig-Maximilians-Universit\"{a}t\\Theresienstr. 37, D-80333 M\"{u}nchen}
\title{Non-relativistic Schr\"{o}dinger theory on q-deformed quantum spaces II\\{\small The free non-relativistic particle and its interactions}}
\date{}
\maketitle

\begin{abstract}
\noindent This is the second part of a paper about a q-deformed analog of
non-relativistic Schr\"{o}dinger theory. It applies the general ideas of part
I and tries to give a description of one-particle states on q-deformed quantum
spaces like the braided line or the q-deformed Euclidean space in three
dimensions. Hamiltonian operators for the free q-deformed particle in one as
well as three dimensions are introduced. Plane waves as solutions to the
corresponding Schr\"{o}dinger equations are considered. Their completeness and
orthonormality relations are written down. Expectation values of position and
momentum observables are taken with respect to one-particle states \ and
their\ time-dependence is discussed. A potential is added to the free-particle
Hamiltonians and q-analogs of the Ehrenfest theorem are derived from the
Heisenberg equations of motion. The conservation of probability is
proved.\newpage

\end{abstract}
\tableofcontents

\section{Introduction}

There is a great hope in physics that lattice-like space-time structures can
help to overcome the difficulties with infinities in quantum field theory
\cite{Schw,Heis}. Snyder's concept of 'quantized space-time' was one of the
first attempts to implement that idea \cite{Sny47, Yan47}. Due to its
attractiveness other researchers took up this idea over and over again
\cite{Fli48, Hill55, Das60, Gol63}. A more recent but very promising approach
to the problem of discretizing space-time arises from the theory of quantum
groups and quantum spaces \cite{Ku83, Wor87, Dri85, Jim85, Drin86, RFT90,
Tak90, WZ91, CSW91, Song92, OSWZ92, Maj93-2, FLW96, Wes97, CW98, GKP96,
MajReg, Oec99, Maj95, KS97, ChDe96, Man88, Maj94-10, Maj93-Int, Wess00,
Maj95star}. If quantum groups and quantum spaces indeed imply a more detailed
description of space-time, they should lead to a mathematical theory\ being
compatible with successful conceptions in physics \cite{MD91, SL92, Fio93,
Blo03, WW01, BW01, Wac02, Wac03, Wac04, Wac05, MSW04, SW04, qAn}.

In part I of this paper we started developing a non-relativistic
Schr\"{o}dinger theory on q-deformed quantum spaces as the braided line or the
q-deformed Euclidean space in three dimensions. Such a program can help to get
a better understanding of the implications of q-deformation in physics, since
the treatment of more realistic space-time structures like the q-deformed
Minkowski space \cite{CSSW90, PW90, SWZ91, Maj91, LWW97} is very awkward (for
other deformations of space-time see Refs. \cite{Lu92, Cas93, Dob94, DFR95,
ChDe95, ChKu04, Koch04}).

From part I of this paper we know that the braided line as well as the
q-deformed Euclidean space in three dimensions can be extended by a time
coordinate. This way we obtain space-time structures in which time behaves
like a commutative and continuous variable, while in space a lattice can be
singled out. This situation is also reflected in the objects of q-analysis and
the time-evolution operators on these spaces. Especially, we saw that
q-analysis leads to discretized versions of classical partial derivatives,
integrals, and so on, whereas the time-evolution operators are of the same
form as their undeformed counterparts. This observation is in complete
accordance with the fact that time is completely decoupled from space. For
this reason, the Schr\"{o}dinger equations and Heisenberg equations of motion
on the quantum spaces under consideration are of the same form as in the
undeformed case.

In part II of our paper we continue the considerations about a
non-relativistic Schr\"{o}dinger\ theory on q-deformed quantum spaces. We
first apply the general formalism developed in part I in order to describe
free non-relativistic one-particle states. In Sec.\thinspace\ref{FreePart} we
introduce free-particle Hamiltonians and show that q-analogs of plane waves
provide a complete and orthonormal set of solutions to the corresponding
Schr\"{o}dinger equations. Then we calculate expectation values for position
and momentum operators taken with respect to free one-particle states and
discuss their time dependence. Section \ref{TheEhr} is devoted to q-analogs of
the theorem of Ehrenfest. In Sec.\thinspace\ref{ConProb} we prove\ that in our
formalism conservation of probability is satisfied. We close our
considerations by a conclusion in Sec.\thinspace\ref{SecCon}. Finally, it
should be noted that we assume the reader to be familiar with the results and
conventions of part I. In this respect, we recommend to have a look at
Sec.\thinspace3.1 or 3.2 of part I.

\section{Free particles on quantum spaces \label{FreePart}}

In this section we would like to study free one-particle states within the
framework developed in part I of our paper. First of all, we have to find a
Hamiltonian suitable for describing a free particle on the q-deformed quantum
spaces under consideration, i.e. braided line and three-dimensional q-deformed
Euclidean space. After that we consider q-analogs of plane waves and show that
they give a complete and orthonormal set of solutions to Schr\"{o}dinger
equations. Finally, we write down expressions for expectation values of
momentum and position observables taken with respect to one-particle wave functions.

\subsection{Free-particle Hamiltonians}

Clearly, the free-particle Hamiltonian should be invariant under translations
and rotations. Thus, a possible choice is given by an element that spans the
one-dimensional eigenspace of the corresponding R-matrix:

\begin{itemize}
\item[(i)] (braided line)%
\begin{equation}
H_{0}\equiv P_{1}P_{1}(2m)^{-1}, \label{Ham1}%
\end{equation}

\item[(ii)] (q-deformed Euclidean space in three dimensions)%
\begin{equation}
H_{0}\equiv g^{AB}P_{B}P_{A}(2m)^{-1}, \label{Ham2}%
\end{equation}

\end{itemize}

\noindent where $g^{AB}$ denotes the quantum metric of the q-deformed
three-dimensional Euclidean space. The constant $m$ stands for a mass
parameter. It is a central and real element of the momentum algebra. One
should also notice that the momentum operators can be expressed by partial
derivatives, as we have $P_{A}=\,$i$\partial_{A}.\ $In this manner, $H_{0}$
becomes an Hermitian operator. (In textbooks on quantum mechanics one usually
finds the convention $P=\,$i$^{-1}\partial_{A}$, but such a choice would make
our formalism more complicated. For this reason, some expressions in this
paper contain an additional minus sign compared to the formulae the reader may
be familiar with.)

As a consequence of their very definition the Hamiltonians in (\ref{Ham1}) and
(\ref{Ham2}) behave like scalars. On these grounds, they commute with momentum
operators, i.e.%
\begin{equation}
\lbrack H_{0},P_{A}]=0. \label{ComHP}%
\end{equation}
Furthermore, we demand that $H_{0}$ inherits the braiding properties from
$\partial_{0}.$ Realizing that $\partial_{0}$ has trivial braiding this
requirement implies for braided products between the mass parameter $m$ and a
function in position or momentum space that

\begin{itemize}
\item[(i)] (braided line)%
\begin{align}
m\odot_{\bar{L}}f(p_{i})  &  =f(q^{2}p_{1},p_{0})\otimes m,\nonumber\\
m\odot_{L}f(p_{i})  &  =f(q^{-2}p_{1},p_{0})\otimes m,\\[0.1in]
m\odot_{\bar{L}}f(x^{i})  &  =f(q^{-2}x^{1},x^{0})\otimes m,\nonumber\\
m\odot_{L}f(x^{i})  &  =f(q^{2}x^{1},x^{0})\otimes m,
\end{align}

\item[(ii)] (q-deformed Euclidean space in three dimensions)%
\begin{align}
m\odot_{\bar{L}}f(p_{i})  &  =f(q^{4}p_{A},p_{0})\otimes m,\nonumber\\
m\odot_{L}f(p_{i})  &  =f(q^{-4}p_{A},p_{0})\otimes m,\\[0.1in]
m\odot_{\bar{L}}f(x^{i})  &  =f(q^{-4}x^{A},x^{0})\otimes m,\nonumber\\
m\odot_{L}f(x^{i})  &  =f(q^{4}x^{A},x^{0})\otimes m,
\end{align}

\end{itemize}

\noindent where the symbols $\odot_{\hspace{-0.01in}\gamma},$ $\gamma
\in\{L,\bar{L},R,\bar{R}\},$ denote the braided products. These braided
products represent realizations of braiding mappings \cite{Wac05}. One should
also notice that we took the convention from part I that capital letters like
$A,$ $B,$ etc. denote indices of space coordinates, i.e., for example,
$x^{i}=(x^{A},x^{0})=(x^{A},t).$

In part I we derived q-analogs of the Schr\"{o}dinger equation. With the
free-particle Hamiltonians they become%
\begin{align}
\text{i}\partial_{0}\overset{t}{\triangleright}\phi(x^{A},t)  &
=H_{0}\overset{x}{\triangleright}\phi(x^{A},t),\nonumber\\
\text{i}\hat{\partial}_{0}\,\overset{t}{\bar{\triangleright}}\,\phi(x^{A},t)
&  =H_{0}\,\overset{x}{\bar{\triangleright}}\,\phi(x^{A},t),
\label{FreParSch1}%
\end{align}
and
\begin{align}
\phi(x^{A},t)\overset{t}{\triangleleft}(\text{i}\hat{\partial}_{0})  &
=\phi(x^{A},t)\overset{x}{\triangleleft}H_{0},\nonumber\\
\phi(x^{A},t)\,\overset{t}{\bar{\triangleleft}}\,(\text{i}\hat{\partial}_{0})
&  =\phi(x^{A},t)\,\overset{x}{\bar{\triangleleft}}\,H_{0}. \label{FreParSch2}%
\end{align}

\subsection{Plane waves}

Its is now our aim to seek solutions to the equations in (\ref{FreParSch1})
and (\ref{FreParSch2}). To this end let us recall that q-exponentials on
quantum spaces play the role of momentum eigenfunctions \cite{Maj93-5, Wac03,
Wac04, qAn, Schir94}. To be more specific, we have%
\begin{align}
\text{i}\partial_{i}\overset{x}{\triangleright}\exp(x^{k}|\text{i}^{-1}%
p_{l})_{\bar{R},L}  &  =\exp(x^{k}|\text{i}^{-1}p_{l})_{\bar{R},L}\overset
{p}{\circledast}p_{i},\nonumber\\
\text{i}\hat{\partial}_{i}\,\overset{x}{\bar{\triangleright}}\,\exp
(x^{k}|\text{i}^{-1}p_{l})_{R,\bar{L}}  &  =\exp(x^{k}|\text{i}^{-1}%
p_{l})_{R,\bar{L}}\overset{p}{\circledast}p_{i},\label{EigGlAnf}\\[0.16in]
\exp(\text{i}^{-1}p_{l}|x^{k})_{\bar{R},L}\,\overset{x}{\bar{\triangleleft}%
}\,(\text{i}\partial^{i})  &  =p_{i}\overset{p}{\circledast}\exp(\text{i}%
^{-1}p_{l}|x^{k})_{\bar{R},L},\nonumber\\
\exp(\text{i}^{-1}p_{l}|x^{k})_{R,\bar{L}}\overset{x}{\triangleleft}%
(\text{i}\hat{\partial}^{i})  &  =p_{i}\overset{p}{\circledast}\exp
(\text{i}^{-1}p_{l}|x^{k})_{R,\bar{L}}. \label{EigGlEnd}%
\end{align}

With these equalities at hand one can prove that solutions to the
Schr\"{o}dinger equations on the braided line are given by the functions
\begin{align}
(u_{\bar{R},L})_{p,m}(x^{i})  &  \equiv\exp(x^{i}|\text{i}^{-1}p_{j})_{\bar
{R},L}\big |_{p_{0}=(2m)^{-1}(p_{1})^{2}}\nonumber\\
&  =\sum_{n_{0},n_{1}=0}^{\infty}\frac{1}{n_{0}![[n_{1}]]_{q}!}(x^{0})^{n_{0}%
}(x^{1})^{n_{1}}\otimes(\text{i}^{-1}p_{1})^{2n_{0}+n_{1}}(2m)^{-n_{0}%
}\nonumber\\
&  =\exp(x^{0}\otimes\text{i}^{-1}(p_{1})^{2}(2m)^{-1})\,\exp_{q}(x^{1}%
\otimes\text{i}^{-1}p_{1}),\label{Sol1dimAnf}\\[0.16in]
(u_{R,\bar{L}})_{p,m}(x^{i})  &  \equiv\exp(x^{i}|\text{i}^{-1}p_{j}%
)_{R,\bar{L}}\big |_{p_{0}=(2m)^{-1}(p_{1})^{2}}\nonumber\\
&  =\sum_{n_{0},n_{1}=0}^{\infty}\frac{1}{n_{0}![[n_{1}]]_{q^{-1}}!}%
(x^{0})^{n_{0}}(x^{1})^{n_{1}}\otimes(\text{i}^{-1}p_{1})^{2n_{0}+n_{1}%
}(2m)^{-n_{0}}\nonumber\\
&  =\exp(x^{0}\otimes\text{i}^{-1}(p_{1})^{2}(2m)^{-1})\,\exp_{q^{-1}}%
(x^{1}\otimes\text{i}^{-1}p_{1}),
\end{align}
and%
\begin{align}
(\bar{u}_{\bar{R},L})_{p,m}(x^{i})  &  \equiv\exp(\text{i}^{-1}p_{j}%
|x^{i})_{\bar{R},L}\big |_{p_{0}=(p_{1})^{2}(2m)^{-1}}\nonumber\\
&  =\sum_{n_{0},n_{1}=0}^{\infty}\frac{1}{n_{0}![[n_{1}]]_{q}!}(2m)^{-n_{0}%
}(-\text{i}^{-1}p_{1})^{2n_{0}+n_{1}}\otimes(x^{0})^{n_{0}}(x^{1})^{n_{1}%
}\nonumber\\
&  =\exp(-\text{i}^{-1}(2m)^{-1}(p_{1})^{2}\otimes x^{0})\,\exp_{q}%
(-\text{i}^{-1}p_{1}\otimes x^{1}),\\[0.16in]
(\bar{u}_{R,\bar{L}})_{p,m}(x^{i})  &  \equiv\exp(\text{i}^{-1}p_{j}%
|x^{i})_{R,\bar{L}}\big |_{p_{0}=(p_{1})^{2}(2m)^{-1}}\nonumber\\
&  =\sum_{n_{0},n_{1}=0}^{\infty}\frac{1}{n_{0}![[n_{1}]]_{q^{-1}}%
!}(2m)^{-n_{0}}(-\text{i}^{-1}p_{1})^{2n_{0}+n_{1}}\otimes(x^{0})^{n_{0}%
}(x^{1})^{n_{1}}\nonumber\\
&  =\exp(-\text{i}^{-1}(2m)^{-1}(p_{1})^{2}\otimes x^{0})\,\exp_{q^{-1}%
}(-\text{i}^{-1}p_{1}\otimes x^{1}). \label{Sol1dimEnd}%
\end{align}
Using the relations in (\ref{EigGlAnf}) and (\ref{EigGlEnd}) together with the
explicit form for the free-particle Hamiltonian one readily checks that%
\begin{align}
\text{i}\partial_{0}\overset{t}{\triangleright}(u_{\bar{R},L})_{p,m}(x^{i})
&  =H_{0}\overset{x}{\triangleright}(u_{\bar{R},L})_{p,m}(x^{i}),\nonumber\\
\text{i}\hat{\partial}_{0}\,\overset{t}{\bar{\triangleright}}\,(u_{R,\bar{L}%
})_{p,m}(x^{i})  &  =H_{0}\,\overset{x}{\bar{\triangleright}}\,(u_{R,\bar{L}%
})_{p,m}(x^{i}), \label{FreSchGlImp1}%
\end{align}
and
\begin{align}
(\bar{u}_{R,\bar{L}})_{p,m}(x^{i})\overset{t}{\triangleleft}(\text{i}%
\hat{\partial}_{0})  &  =(\bar{u}_{R,\bar{L}})_{p,m}(x^{i})\overset
{x}{\triangleleft}H_{0},\nonumber\\
(\bar{u}_{\bar{R},L})_{p,m}(x^{i})\,\overset{t}{\bar{\triangleleft}%
}\,(\text{i}\hat{\partial}_{0})  &  =(\bar{u}_{\bar{R},L})_{p,m}%
(x^{i})\,\overset{x}{\bar{\triangleleft}}\,H_{0}. \label{FreSchGlImp4}%
\end{align}

Now, we come to the solutions for the three-dimensional q-deformed Euclidean
space. In this case, however, we have to work a little bit harder. Again, we
substitute $p^{2}(2m)^{-1}=g^{AB}p_{B}\circledast p_{A}(2m)^{-1}$ for $p_{0}$,
but now we have to apply star multiplication \cite{WW01, BFF78, Moy49,
MSSW00}. In this manner, the unconjugate solutions to the three-dimensional
Schr\"{o}dinger equations then become%
\begin{align}
&  (u_{\bar{R},L})_{p,m}(x^{i})\equiv\exp(x^{i}|\text{i}^{-1}p_{j})_{\bar
{R},L}\big |_{p_{0}=\,(p_{1})^{2}(2m)^{-1}\circledast}\nonumber\\
&  \qquad=\sum_{\underline{n}=0}^{\infty}\sum_{k=0}^{n_{0}-1}\frac
{(\text{i}^{-1})^{n_{0}}(-\lambda_{+})^{n_{0}-k}q^{-2k+2n_{3}(n_{0}-k)}}%
{n_{0}![[n_{+}]]_{q^{4}}![[n_{3}]]_{q^{2}}![[n_{-}]]_{q^{4}}!}\,%
%TCIMACRO{\QATOPD{[}{]}{n_{0}}{k}}%
%BeginExpansion
\genfrac{[}{]}{0pt}{}{n_{0}}{k}%
%EndExpansion
_{q^{4}}\nonumber\\
&  \qquad\qquad\quad\times(x^{0})^{n_{0}}(x^{+})^{n_{+}}(x^{3})^{n_{3}}%
(x^{-})^{n_{-}}\nonumber\\
&  \qquad\qquad\quad\otimes(\text{i}^{-1}p_{-})^{n_{-}+\,n_{0}-k}%
(\text{i}^{-1}p_{3})^{n_{3}+2k}(\text{i}^{-1}p_{+})^{n_{+}+n_{0}%
-k}(2m)^{-n_{0}},\label{Sol3dimAnf}\\[0.1in]
&  (u_{R,\bar{L}})_{p,m}(x^{i})\equiv\exp(x^{i}|\text{i}^{-1}p_{j})_{R,\bar
{L}}\big |_{p_{0}=\,(p_{1})^{2}(2m)^{-1}\circledast}\nonumber\\
&  \qquad=\sum_{\underline{n}=0}^{\infty}\sum_{k=0}^{n_{0}-1}\frac
{(\text{i}^{-1})^{n_{0}}(-\lambda_{+})^{n_{0}-k}q^{2k-2n_{3}(n_{0}-k)}}%
{n_{0}![[n_{+}]]_{q^{-4}}![[n_{3}]]_{q^{-2}}![[n_{-}]]_{q^{-4}}!}\,%
%TCIMACRO{\QATOPD{[}{]}{n_{0}}{k}}%
%BeginExpansion
\genfrac{[}{]}{0pt}{}{n_{0}}{k}%
%EndExpansion
_{q^{-4}}\nonumber\\
&  \qquad\qquad\quad\times(x^{0})^{n_{0}}(x^{-})^{n_{-}}(x^{3})^{n_{3}}%
(x^{+})^{n_{+}}\nonumber\\
&  \qquad\qquad\quad\otimes(\text{i}^{-1}p_{+})^{n_{+}+\,n_{0}-k}%
(\text{i}^{-1}p_{3})^{n_{3}+2k}(\text{i}^{-1}p_{-})^{n_{-}+\,n_{0}%
-k}(2m)^{-n_{0}}, \label{Sol3dimEnd}%
\end{align}
and for the conjugate solutions we likewise find%
\begin{align}
&  (\bar{u}_{\bar{R},L})_{p,m}(x^{i})\equiv\exp(\text{i}^{-1}p_{j}%
|x^{i})_{\bar{R},L}\big |_{p_{0}=\,(2m)^{-1}p^{2}\overset{p}{\circledast}%
}\nonumber\\
&  \qquad=\sum_{\underline{n}=0}^{\infty}\sum_{k=0}^{n_{0}-1}\frac
{(-\text{i})^{n_{0}}(-\lambda_{+})^{n_{0}-k}q^{-2k+2n_{3}(n_{0}-k)}}%
{n_{0}![[n_{+}]]_{q^{4}}![[n_{3}]]_{q^{2}}![[n_{-}]]_{q^{4}}!}\,%
%TCIMACRO{\QATOPD{[}{]}{n_{0}}{k}}%
%BeginExpansion
\genfrac{[}{]}{0pt}{}{n_{0}}{k}%
%EndExpansion
_{q^{4}}\nonumber\\
&  \qquad\qquad\quad\times(2m)^{-n_{0}}(-\text{i}^{-1}p_{-})^{n_{-}+\,n_{0}%
-k}(-\text{i}^{-1}p_{3})^{n_{3}+2k}(-\text{i}^{-1}p_{+})^{n_{+}+\,n_{0}%
-k}\nonumber\\
&  \qquad\qquad\quad\otimes(x^{0})^{n_{0}}(x^{+})^{n_{+}}(x^{3})^{n_{3}}%
(x^{-})^{n_{-}},\\[0.1in]
&  (\bar{u}_{R,\bar{L}})_{p,m}(x^{i})\equiv\exp(\text{i}^{-1}p_{j}%
|x^{i})_{R,\bar{L}}\big |_{p_{0}=\,(2m)^{-1}p^{2}\overset{p}{\circledast}%
}\nonumber\\
&  \qquad=\sum_{\underline{n}=0}^{\infty}\sum_{k=0}^{n_{0}-1}\frac
{(-\text{i})^{n_{0}}(-\lambda_{+})^{n_{0}-k}q^{2k-2n_{3}(n_{0}-k)}}%
{n_{0}![[n_{+}]]_{q^{-4}}![[n_{3}]]_{q^{-2}}![[n_{-}]]_{q^{-4}}!}\,%
%TCIMACRO{\QATOPD{[}{]}{n_{0}}{k}}%
%BeginExpansion
\genfrac{[}{]}{0pt}{}{n_{0}}{k}%
%EndExpansion
_{q^{-4}}\nonumber\\
&  \qquad\qquad\quad\times(2m)^{-n_{0}}(-\text{i}^{-1}p_{+})^{n_{+}+\,n_{0}%
-k}(-\text{i}^{-1}p_{3})^{n_{3}+2k}(-\text{i}^{-1}p_{-})^{n_{-}+\,n_{0}%
-k}\nonumber\\
&  \qquad\qquad\quad\otimes(x^{0})^{n_{0}}(x^{-})^{n_{-}}(x^{3})^{n_{3}}%
(x^{+})^{n_{+}}, \label{Sol3dimEndN}%
\end{align}
where $\lambda_{+}=q+q^{-1}.$ The q-binomial coefficients are defined by the
formula \cite{KS97}
\begin{equation}%
%TCIMACRO{\QATOPD{[}{]}{\alpha}{k}}%
%BeginExpansion
\genfrac{[}{]}{0pt}{}{\alpha}{k}%
%EndExpansion
_{q^{a}}\equiv\frac{\left[  \left[  \alpha\right]  \right]  _{q^{a}}\left[
\left[  \alpha-1\right]  \right]  _{q^{a}}\ldots\left[  \left[  \alpha
-k+1\right]  \right]  _{q^{a}}}{\left[  \left[  k\right]  \right]  _{q^{a}}!},
\end{equation}
with $\alpha\in\mathbb{C},$ $k\in\mathbb{N}$.

We would like to say a few words about the ideas the derivation of the
expressions in (\ref{Sol3dimAnf})-(\ref{Sol3dimEndN}) is based on. We
concentrate attention to the expression in (\ref{Sol3dimAnf}), since the other
formulae follow from similar reasonings. First of all we make as ansatz
\begin{equation}
\underset{n\text{-times}}{\underbrace{p^{2}\overset{p}{\circledast}%
\ldots\overset{p}{\circledast}p^{2}}}=\sum_{k=0}^{n}(C_{q})_{k}^{n}%
\,(p_{-})^{n-k}(p_{3})^{2k}(p_{+})^{n-k}.
\end{equation}
Exploiting the commutation relations of three-dimensional q-deformed Euclidean
space \cite{LWW97} we find that the coefficients $(C_{q})_{k}^{n}$ are subject
to the recursion relation%
\begin{equation}
(C_{q})_{k}^{n}=q^{4k}(-\lambda_{+})(C_{q})_{k}^{n-1}+q^{-2}(C_{q}%
)_{k-1}^{n-1}.
\end{equation}
As one can prove by inserting, the above recursion relation has the solution%
\begin{equation}
(C_{q})_{k}^{n}=q^{-2k}(-\lambda_{+})^{n-k}%
%TCIMACRO{\QATOPD{[}{]}{n}{k}}%
%BeginExpansion
\genfrac{[}{]}{0pt}{}{n}{k}%
%EndExpansion
_{q^{4}}.
\end{equation}
From what we have done so far we get%
\begin{align}
&  \underset{n_{0}\text{-times}}{\underbrace{p^{2}\overset{p}{\circledast
}\ldots\overset{p}{\circledast}p^{2}}}\overset{p}{\circledast}(p_{-})^{n_{-}%
}(p_{3})^{n_{3}}(p_{+})^{n_{+}}=\nonumber\\
&  \qquad=\,(p_{-})^{n_{-}}\underset{n_{0}\text{-times}}{\overset
{p}{\circledast}\underbrace{p^{2}\overset{p}{\circledast}\ldots\overset
{p}{\circledast}p^{2}}}\overset{p}{\circledast}(p_{3})^{n_{3}}(p_{+})^{n_{+}%
}\nonumber\\
&  \qquad=\,\sum_{k=0}^{n_{0}}(C_{q})_{k}^{n}\,(p_{-})^{n_{-}+\,n_{0}-k}%
(p_{3})^{2k}(p_{+})^{n_{0}-k}\overset{p}{\circledast}(p_{3})^{n_{3}}%
(p_{+})^{n_{+}}\nonumber\\
&  \qquad=\,\sum_{k=0}^{n_{0}}q^{-2k}(C_{q})_{k}^{n}\,(p_{-})^{n_{-}%
+\,n_{0}-k}(p_{3})^{n_{3}+2k}(p_{3})^{n_{3}}(p_{+})^{n_{+}+\,n_{0}-k}.
\end{align}
The point now is that the function $(u_{\bar{R},L})_{p,m}(x^{i})$ arises from
the q-exponential $\exp(x^{i}|$i$^{-1}p_{j})_{\bar{R},L}$ by applying the
substitution%
\begin{equation}
(p^{0})^{n_{0}}(p_{-})^{n_{-}}(p_{3})^{n_{3}}(p_{+})^{n_{+}}\rightarrow
\underset{n_{0}\text{-times}}{\underbrace{\frac{p^{2}}{2m}\overset
{p}{\circledast}\ldots\overset{p}{\circledast}\frac{p^{2}}{2m}}}\overset
{p}{\circledast}(p_{-})^{n_{-}}(p_{3})^{n_{3}}(p_{+})^{n_{+}}.
\end{equation}
These arguments finally lead us to the last expression in (\ref{Sol3dimAnf}).

Sometimes it is convenient to write the functions in (\ref{Sol3dimAnf}%
)-(\ref{Sol3dimEndN}) in a way that makes their dependence from time more
explicit. In this manner we have%
\begin{align}
(u_{\bar{R},L})_{p,m}(x^{i})  &  =\exp(x^{i}|\text{i}^{-1}p_{j})_{\bar{R}%
,L}\big |_{x^{0}=0}\overset{p}{\circledast}\exp(-\text{i}tp^{2}(2m)^{-1}%
)_{\bar{R},L},\nonumber\\
(u_{R,\bar{L}})_{p,m}(x^{i})  &  =\exp(x^{i}|\text{i}^{-1}p_{j})_{R,\bar{L}%
}\big |_{x^{0}=0}\overset{p}{\circledast}\exp(-\text{i}tp^{2}(2m)^{-1}%
)_{R,\bar{L}},\label{ExpTimDep1}\\[0.1in]
(\bar{u}_{\bar{R},L})_{p,m}(x^{i})  &  =\exp(\text{i}(2m)^{-1}p^{2}t)_{\bar
{R},L}\overset{p}{\circledast}\big (\exp(\text{i}^{-1}p_{j}|x^{i})_{\bar{R}%
,L}\big )\big |_{x^{0}=0},\nonumber\\
(\bar{u}_{R,\bar{L}})_{p,m}(x^{i})  &  =\exp(\text{i}(2m)^{-1}p^{2}%
t)_{R,\bar{L}}\overset{p}{\circledast}\big (\exp(\text{i}^{-1}p_{j}%
|x^{i})_{R,\bar{L}}\big )\big |_{x^{0}=0}, \label{ExpTimDep2}%
\end{align}
where the time-dependent phase factors take the form

\begin{itemize}
\item[(i)] (braided line)%
\begin{align}
\exp(-\text{i}tp^{2}(2m)^{-1})_{\bar{R},L}  &  =\exp(-\text{i}tp^{2}%
(2m)^{-1})_{R,\bar{L}}\nonumber\\
&  =\sum_{n=0}^{\infty}\frac{1}{n!}\left(  -\text{i}tp_{1}\right)
^{n}(2m)^{-n},\\[0.1in]
\exp(\text{i}(2m)^{-1}p^{2}t)_{\bar{R},L}  &  =\exp(\text{i}(2m)^{-1}%
p^{2}t)_{R,\bar{L}}\nonumber\\
&  =\sum_{n=0}^{\infty}\frac{1}{n!}(2m)^{-n}\left(  \text{i}p_{1}t\right)
^{n},
\end{align}

\item[(ii)] (q-deformed Euclidean space in three dimensions)%
\begin{align}
\exp(-\text{i}tp^{2}(2m)^{-1})_{\bar{R},L}  &  =\sum_{n=0}^{\infty}%
\frac{(-\text{i}t)^{n}}{n!}\sum_{k=0}^{n}q^{-2k}(-\lambda_{+})^{n-k}%
%TCIMACRO{\QATOPD{[}{]}{n}{k}}%
%BeginExpansion
\genfrac{[}{]}{0pt}{}{n}{k}%
%EndExpansion
_{q^{4}}\nonumber\\
&  \qquad\times(p_{-})^{n-k}(p_{3})^{2k}(p_{+})^{n-k}(2m)^{-n},\\[0.1in]
\exp(-\text{i}tp^{2}(2m)^{-1})_{R,\bar{L}}  &  =\sum_{n=0}^{\infty}%
\frac{(-\text{i}t)^{n}}{n!}\sum_{k=0}^{n}q^{2k}(-\lambda_{+})^{n-k}%
%TCIMACRO{\QATOPD{[}{]}{n}{k}}%
%BeginExpansion
\genfrac{[}{]}{0pt}{}{n}{k}%
%EndExpansion
_{q^{-4}}\nonumber\\
&  \qquad\times(p_{+})^{n-k}(p_{3})^{2k}(p_{-})^{n-k}(2m)^{-n},\\[0.1in]
\exp(\text{i}(2m)^{-1}p^{2}t)_{\bar{R},L}  &  =\sum_{n=0}^{\infty}%
\frac{(\text{i}t)^{n}}{n!}\sum_{k=0}^{n}q^{-2k}(-\lambda_{+})^{n-k}%
%TCIMACRO{\QATOPD{[}{]}{n}{k}}%
%BeginExpansion
\genfrac{[}{]}{0pt}{}{n}{k}%
%EndExpansion
_{q^{4}}\nonumber\\
&  \qquad\times(2m)^{-n}(p_{-})^{n-k}(p_{3})^{2k}(p_{+})^{n-k},\\[0.1in]
\exp(\text{i}(2m)^{-1}p^{2}t)_{R,\bar{L}}  &  =\sum_{n=0}^{\infty}%
\frac{(\text{i}t)^{n}}{n!}\sum_{k=0}^{n}q^{2k}(-\lambda_{+})^{n-k}%
%TCIMACRO{\QATOPD{[}{]}{n}{k}}%
%BeginExpansion
\genfrac{[}{]}{0pt}{}{n}{k}%
%EndExpansion
_{q^{-4}}\nonumber\\
&  \qquad\times(2m)^{-n}(p_{+})^{n-k}(p_{3})^{2k}(p_{-})^{n-k}.
\end{align}

\end{itemize}

\noindent In the case of the braided line the star product in the relations of
(\ref{ExpTimDep1}) and (\ref{ExpTimDep2}) is given by the commutative product.
For later purpose we would like to mention that the conjugate solutions
$(\bar{u}_{\bar{R},L})_{p,m}$ and $(\bar{u}_{R,\bar{L}})_{p,m}$ describe
particles traversing backwards in time, as can be seen from the equations in
(\ref{ExpTimDep1}) and (\ref{ExpTimDep2}).

To sum up, we found q-analogs of the stationary solutions to the
Schr\"{o}dinger equation of a free non-relativistic particle. In analogy to
the undeformed case they are eigenfunctions of energy and momentum, since we
have%
\begin{align}
P_{A}\overset{x}{\triangleright}(u_{\bar{R},L})_{p,m}(x^{i})  &
=\text{i}\partial_{A}\overset{x}{\triangleright}(u_{\bar{R},L})_{p,m}%
(x^{i})=(u_{\bar{R},L})_{p,m}(x^{i})\overset{p}{\circledast}p_{A},\nonumber\\
H_{0}\overset{x}{\triangleright}(u_{\bar{R},L})_{p,m}(x^{i})  &
=(2m)^{-1}P^{2}\overset{x}{\triangleright}(u_{\bar{R},L})_{p,m}(x^{i}%
)\nonumber\\
&  =(u_{\bar{R},L})_{p,m}(x^{i})\overset{p}{\circledast}p^{2}(2m)^{-1}%
,\label{EigGle1}\\[0.1in]
P_{A}\,\overset{x}{\bar{\triangleright}}\,(u_{R,\bar{L}})_{p,m}(x^{i})  &
=\text{i}\hat{\partial}_{A}\,\overset{x}{\bar{\triangleright}}\,(u_{R,\bar{L}%
})_{p,m}(x^{i})=(u_{R,\bar{L}})_{p,m}(x^{i})\overset{p}{\circledast}%
p_{A},\nonumber\\
H_{0}\,\overset{x}{\bar{\triangleright}}\,(u_{R,\bar{L}})_{p,m}(x^{i})  &
=(2m)^{-1}P^{2}\,\overset{x}{\bar{\triangleright}}\,(u_{R,\bar{L}}%
)_{p,m}(x^{i})\nonumber\\
&  =(u_{R,\bar{L}})_{p,m}(x^{i})\overset{p}{\circledast}p^{2}(2m)^{-1},
\end{align}
and%
\begin{align}
(\bar{u}_{\bar{R},L})_{p,m}(x^{i})\,\overset{x}{\bar{\triangleleft}}\,P_{A}
&  =(\bar{u}_{\bar{R},L})_{p,m}(x^{i})\,\overset{x}{\bar{\triangleleft}%
}\,\text{i}\partial_{A}=p_{A}\overset{p}{\circledast}(\bar{u}_{\bar{R}%
,L})_{p,m}(x^{i}),\nonumber\\
(\bar{u}_{\bar{R},L})_{p,m}(x^{i})\,\overset{x}{\bar{\triangleleft}}\,H_{0}
&  =(\bar{u}_{\bar{R},L})_{p,m}(x^{i})\,\overset{x}{\bar{\triangleleft}%
}\,P^{2}(2m)^{-1}\nonumber\\
&  =(2m)^{-1}p^{2}\overset{p}{\circledast}(\bar{u}_{\bar{R},L})_{p,m}%
(x^{i}),\\[0.1in]
(\bar{u}_{R,\bar{L}})_{p,m}(x^{i})\overset{x}{\triangleleft}P_{A}  &
=(\bar{u}_{R,\bar{L}})_{p,m}(x^{i})\overset{x}{\triangleleft}\text{i}%
\hat{\partial}_{A}=p_{A}\overset{p}{\circledast}(\bar{u}_{R,\bar{L}}%
)_{p,m}(x^{i}),\nonumber\\
(\bar{u}_{R,\bar{L}})_{p,m}(x^{i})\overset{x}{\triangleleft}H_{0}  &
=(\bar{u}_{R,\bar{L}})_{p,m}(x^{i})\overset{x}{\triangleleft}P^{2}%
(2m)^{-1}\nonumber\\
&  =(2m)^{-1}p^{2}\overset{p}{\circledast}(\bar{u}_{R,\bar{L}})_{p,m}(x^{i}).
\label{EigGle4}%
\end{align}
These relations are in accordance with the observation that energy and
momentum commute with each other [cf. Eq. (\ref{ComHP})].

In part I of this paper we discussed how the time evolution operators look
like on the quantum spaces under consideration. For the sake of completeness
it should be noted that our solutions can alternatively be obtained by
applying these time evolution operators onto time-independent plane waves,
i.e.%
\begin{align}
(u_{\bar{R},L})_{p,m}(x^{i})  &  =\exp(-t\otimes\text{i}H_{0})\overset
{H_{0}|x}{\triangleright}\big (\exp(x^{i}|\text{i}^{-1}p_{j})_{\bar{R}%
,L}\big |_{x^{0}=0}\big ),\nonumber\\
(u_{R,\bar{L}})_{p,m}(x^{i})  &  =\exp(-t\otimes\text{i}H_{0})\,\overset
{H_{0}|x}{\bar{\triangleright}}\,\big (\exp(x^{i}|\text{i}^{-1}p_{j}%
)_{R,\bar{L}}\big |_{x^{0}=0}\big ),
\end{align}
and%
\begin{align}
(\bar{u}_{\bar{R},L})_{p,m}(x^{i})  &  =\exp(\text{i}^{-1}p_{l}|x^{k}%
)_{\bar{R},L}\big |_{x^{0}=0}\,\overset{x|H_{0}}{\bar{\triangleleft}}%
\,\exp(\text{i}H_{0}\otimes t),\nonumber\\
(\bar{u}_{R,\bar{L}})_{p,m}(x^{i})  &  =\exp(\text{i}^{-1}p_{l}|x^{k}%
)_{R,\bar{L}}\big |_{x^{0}=0}\overset{x|H_{0}}{\triangleleft}\exp
(\text{i}H_{0}\otimes t). \label{ConPlaWav}%
\end{align}

Before we proceed any further let us note that the functions $(u_{\bar{R}%
,L})_{\ominus_{L}p,m},$ $(u_{R,\bar{L}})_{\ominus_{\bar{L}}p,m},$ $(\bar
{u}_{\bar{R},L})_{\ominus_{\bar{R}}p,m},$ and $(\bar{u}_{R,\bar{L}}%
)_{\ominus_{R}p,m}$ give further solutions to the free-particle
Schr\"{o}dinger equations. (Notice that the operations $\ominus_{R}$ and
$\ominus_{\bar{R}}$ can be viewed as right versions of $\ominus_{L}$ and
$\ominus_{\bar{L}}$, respectively. We did not mention them explicitly in part
I, since we can make the identifications $f(\ominus_{R}\,x^{i})=f(\ominus
_{\bar{L}}\,x^{i}),\ f(\ominus_{\bar{R}}x^{i})=f(\ominus_{L}\,x^{i})$. Similar
reasonings hold for the operations $\oplus_{\gamma}$ and $\odot_{\gamma},$
$\gamma\in\{L,\bar{L},R,\bar{R}\}.$) Applying the operations $\ominus_{\gamma
}$ to the momentum part of the equations in (\ref{ExpTimDep1}) and
(\ref{ExpTimDep2}) one readily checks that
\begin{align}
\text{i}\partial_{0}\overset{t}{\triangleright}(u_{\bar{R},L})_{\ominus
_{L}p,m}(x^{i})  &  =H_{0}\overset{x}{\triangleright}(u_{\bar{R},L}%
)_{\ominus_{L}p,m}(x^{i}),\nonumber\\
\text{i}\hat{\partial}_{0}\,\overset{t}{\bar{\triangleright}}\,(u_{R,\bar{L}%
})_{\ominus_{\bar{L}}p,m}(x^{i})  &  =H_{0}\,\overset{x}{\bar{\triangleright}%
}\,(u_{R,\bar{L}})_{\ominus_{\bar{L}}p,m}(x^{i}), \label{UnConSol}%
\end{align}
and
\begin{align}
(\bar{u}_{R,\bar{L}})_{\ominus_{R}p,m}(x^{i})\overset{t}{\triangleleft
}(\text{i}\hat{\partial}_{0})  &  =(\bar{u}_{R,\bar{L}})_{\ominus_{R}%
p,m}(x^{i})\overset{x}{\triangleleft}H_{0},\nonumber\\
(\bar{u}_{\bar{R},L})_{\ominus_{\bar{R}}p,m}(x^{i})\,\overset{t}%
{\bar{\triangleleft}}\,(\text{i}\hat{\partial}_{0})  &  =(\bar{u}_{\bar{R}%
,L})_{\ominus_{\bar{R}}p,m}(x^{i})\,\overset{x}{\bar{\triangleleft}}\,H_{0}.
\label{ConSol}%
\end{align}

Again, the unconjugate solutions in (\ref{UnConSol}) move forward in time,
while the conjugate ones in (\ref{ConSol}) move oppositely. For a better
understanding of the new solutions the reader should be aware of the relations%
\begin{align}
(u_{\bar{R},L})_{\ominus_{L}p,m}(x^{i})  &  =(u_{\bar{R},L})_{p,m}%
(\ominus_{\bar{R}}\,x^{A},t)\nonumber\\
&  \neq(u_{\bar{R},L})_{p,m}(\ominus_{\bar{R}}\,x^{i})=(u_{\bar{R},L}%
)_{p,m}(\ominus_{\bar{R}}\,x^{A},-t),\\[0.1in]
(u_{R,\bar{L}})_{\ominus_{\bar{L}}p,m}(x^{i})  &  =(u_{R,\bar{L}}%
)_{p,m}(\ominus_{R}\,x^{A},t)\nonumber\\
&  \neq(u_{R,\bar{L}})_{p,m}(\ominus_{R}\,x^{i})=(u_{R,\bar{L}})_{p,m}%
(\ominus_{R}\,x^{A},-t),
\end{align}
and%
\begin{align}
(\bar{u}_{\bar{R},L})_{\ominus_{\bar{R}}p,m}(x^{i})  &  =(\bar{u}_{\bar{R}%
,L})_{p,m}(\ominus_{L}\,x^{A},t)\nonumber\\
&  \neq(\bar{u}_{\bar{R},L})_{p,m}(\ominus_{L}\,x^{i})=(u_{\bar{R},L}%
)_{p,m}(\ominus_{L}\,x^{A},-t),\\[0.1in]
(\bar{u}_{R,\bar{L}})_{\ominus_{R}p,m}(x^{i})  &  =(\bar{u}_{R,\bar{L}}%
)_{p,m}(\ominus_{\bar{L}}\,x^{A},t)\nonumber\\
&  \neq(\bar{u}_{R,\bar{L}})_{p,m}(\ominus_{\bar{L}}\,x^{i})=(\bar{u}%
_{R,\bar{L}})_{p,m}(\ominus_{\bar{L}}\,x^{A},-t).
\end{align}

From now on we call this second set of solutions to the Schr\"{o}dinger
equations inverse momentum eigenfunctions. In what follows they will play an
important role, so we would like to discuss their properties further. First of
all, they are again eigenfunctions of energy. Concretely, we have%
\begin{align}
H_{0}\overset{x}{\triangleright}(u_{\bar{R},L})_{\ominus_{L}p,m}(x^{i})  &
=(2m)^{-1}P^{2}\overset{x}{\triangleright}(u_{\bar{R},L})_{\ominus_{L}%
p,m}(x^{i})\nonumber\\
&  =q^{\zeta}(u_{\bar{R},L})_{\ominus_{L}p,m}(x^{i})\overset{p}{\circledast
}p^{2}(2m)^{-1},\\
H_{0}\,\overset{x}{\bar{\triangleright}}\,(u_{R,\bar{L}})_{\ominus_{\bar{L}%
}p,m}(x^{i})  &  =(2m)^{-1}P^{2}\,\overset{x}{\bar{\triangleright}%
}\,(u_{R,\bar{L}})_{\ominus_{\bar{L}}p,m}(x^{i})\nonumber\\
&  =q^{-\zeta}(u_{R,\bar{L}})_{\ominus_{\bar{L}}p,m}(x^{i})\overset
{p}{\circledast}p^{2}(2m)^{-1},
\end{align}
and%
\begin{align}
(\bar{u}_{\bar{R},L})_{\ominus_{\bar{R}}p,m}(x^{i})\,\overset{x}%
{\bar{\triangleleft}}\,H_{0}  &  =(\bar{u}_{\bar{R},L})_{\ominus_{\bar{R}}%
p,m}(x^{i})\,\overset{x}{\bar{\triangleleft}}\,(2m)^{-1}P^{2}\nonumber\\
&  =q^{\zeta}(2m)^{-1}p^{2}\overset{p}{\circledast}(\bar{u}_{\bar{R}%
,L})_{\ominus_{\bar{R}}p,m}(x^{i}),\\
(\bar{u}_{R,\bar{L}})_{\ominus_{R}p,m}(x^{i})\overset{x}{\triangleleft}H_{0}
&  =(\bar{u}_{R,\bar{L}})_{\ominus_{R}p,m}(x^{i})\overset{x}{\triangleleft
}(2m)^{-1}P^{2}\nonumber\\
&  =q^{-\zeta}(2m)^{-1}p^{2}\overset{p}{\circledast}(\bar{u}_{R,\bar{L}%
})_{\ominus_{R}p,m}(x^{i}),
\end{align}
where

\begin{enumerate}
\item[(i)] (braided line) $\zeta=-1,$

\item[(ii)] (q-deformed Euclidean space in three dimensions) $\zeta=2.$
\end{enumerate}

Compared to the relations in (\ref{EigGle1})-(\ref{EigGle4}) the eigenvalues
of energy now contain additional factors. Their occurrence is a consequence of
the fact that we apply the operations $\ominus_{\gamma},$ $\gamma\in
\{L,\bar{L},R,\bar{R}\}$ to momentum eigenfunctions. The concrete form of the
additional factors should become clear from the definition of the operations
$\ominus_{\gamma}$ together with the relations (for notation and conventions
see part I)%
\begin{align}
\ominus_{L}\,p^{2}  &  =(\mathcal{W}_{R}^{-1}\circ S_{L})(P^{2})=q^{\zeta
}p^{2},\nonumber\\
\ominus_{R}\,p^{2}  &  =(\mathcal{W}_{R}^{-1}\circ S_{\bar{L}})(P^{2}%
)=q^{-\zeta}p^{2}.
\end{align}

For the same reasons the time-dependence of inverse momentum eigenfunctions
now takes on the form%
\begin{align}
(u_{\bar{R},L})_{\ominus_{L}p,m}(x^{i})  &  =(u_{\bar{R},L})_{\ominus_{L}%
p,m}(x^{i})\big |_{x^{0}=0}\overset{p}{\circledast}\exp(-\text{i}q^{\zeta
}tp^{2}(2m)^{-1})_{\bar{R},L},\nonumber\\
(u_{R,\bar{L}})_{\ominus_{\bar{L}}p,m}(x^{i})  &  =(u_{R,\bar{L}}%
)_{\ominus_{\bar{L}}p,m}(x^{i})\big |_{x^{0}=0}\overset{p}{\circledast}%
\exp(-\text{i}q^{-\zeta}tp^{2}(2m)^{-1})_{R,\bar{L}}, \label{ExpTimDep3}%
\\[0.1in]
(\bar{u}_{\bar{R},L})_{\ominus_{\bar{R}}p,m}(x^{i})  &  =\exp(\text{i}%
q^{\zeta}(2m)^{-1}p^{2}t)_{\bar{R},L}\overset{p}{\circledast}\big ((\bar
{u}_{\bar{R},L})_{\ominus_{\bar{R}}p,m}(x^{i})\big )\big |_{x^{0}%
=0},\nonumber\\
(\bar{u}_{R,\bar{L}})_{\ominus_{R}p,m}(x^{i})  &  =\exp(\text{i}q^{-\zeta
}(2m)^{-1}p^{2}t)_{R,\bar{L}}\overset{p}{\circledast}\big ((\bar{u}_{R,\bar
{L}})_{\ominus_{R}p,m}(x^{i})\big )\big |_{x^{0}=0}, \label{ExpTimDep4}%
\end{align}
On the other hand we still have the identities%
\begin{align}
(u_{\bar{R},L})_{\ominus_{L}p,m}(x^{i})  &  =\exp(-t\otimes\text{i}%
H_{0})\overset{H_{0}|x}{\triangleright}\big ((u_{\bar{R},L})_{\ominus_{L}%
p,m}(x^{i})\big |_{x^{0}=0}\big ),\nonumber\\
(u_{R,\bar{L}})_{\ominus_{\bar{L}}p,m}(x^{i})  &  =\exp(-t\otimes\text{i}%
H_{0})\,\overset{H_{0}|x}{\bar{\triangleright}}\,\big ((u_{R,\bar{L}%
})_{\ominus_{\bar{L}}p,m}(x^{i})\big |_{x^{0}=0}\big ),
\end{align}
and%
\begin{align}
(\bar{u}_{\bar{R},L})_{\ominus_{\bar{R}}p,m}(x^{i})  &  =(\bar{u}_{\bar{R}%
,L})_{\ominus_{\bar{R}}p,m}(x^{i})\big |_{x^{0}=0}\,\overset{x|H_{0}}%
{\bar{\triangleleft}}\,\exp(\text{i}H_{0}\otimes t),\nonumber\\
(\bar{u}_{R,\bar{L}})_{\ominus_{R}p,m}(x^{i})  &  =(\bar{u}_{R,\bar{L}%
})_{\ominus_{R}p,m}(x^{i})\big |_{x^{0}=0}\overset{x|H_{0}}{\triangleleft}%
\exp(\text{i}H_{0}\otimes t),
\end{align}
since inverse momentum eigenfunctions fulfill the same Schr\"{o}dinger
equations as the momentum eigenfunctions in (\ref{Sol1dimAnf}%
)-(\ref{Sol1dimEnd}) and (\ref{Sol3dimAnf})-(\ref{Sol3dimEnd}).

\subsection{Completeness and orthonormality}

Now, we have everything together to show that in complete analogy to the
undeformed case momentum eigenfunctions on q-deformed quantum spaces establish
a complete and orthonormal set of solutions to the free-particle
Schr\"{o}dinger equations in (\ref{FreSchGlImp1})-(\ref{FreSchGlImp4}). This
observation is a direct consequence of the results in Refs. \cite{Qkin1,
Qkin2}, where we already derived orthonormality and completeness relations for
q-analogs of\ plane waves. Although these plane waves did not satisfy any
energy-momentum relation it is straightforward to adapt the ideas of Refs.
\cite{Qkin1, Qkin2} to our solutions.

Before we explain how to achieve this, let us first write down the explicit
form of the orthonormality and completeness relations, as they read for our
q-deformed momentum eigenfunctions. If we change the normalization of
q-deformed momentum eigenfunctions according to%
\begin{align}
(u_{\bar{R},L})_{p,m}(x^{i})  &  =(\text{vol}_{1})^{-1/2}\exp(x^{i}%
|\text{i}^{-1}p_{j})_{\bar{R},L}\big |_{p_{0}=p^{2}/(2m)\overset
{p}{\circledast}},\nonumber\\
(u_{R,\bar{L}})_{p,m}(x^{i})  &  =(\text{vol}_{2})^{-1/2}\exp(x^{i}%
|\text{i}^{-1}p_{j})_{R,\bar{L}}\big |_{p_{0}=p^{2}/(2m)\overset
{p}{\circledast}},\\[0.1in]
(\bar{u}_{\bar{R},L})_{p,m}(x^{i})  &  =(\text{vol}_{1})^{-1/2}\exp
(\text{i}^{-1}p_{j}|x^{i})_{\bar{R},L}\big |_{p_{0}=p^{2}/(2m)\overset
{p}{\circledast}},\nonumber\\
(\bar{u}_{R,\bar{L}})_{p,m}(x^{i})  &  =(\text{vol}_{2})^{-1/2}\exp
(\text{i}^{-1}p_{j}|x^{i})_{R,\bar{L}}\big |_{p_{0}=p^{2}/(2m)\overset
{p}{\circledast}},
\end{align}
where
\begin{align}
\text{vol}_{1}  &  \equiv\int_{-\infty}^{+\infty}d_{1}^{n}x\int_{-\infty
}^{+\infty}d_{1}^{n}p\exp(x^{i}|\text{i}^{-1}p_{j})_{\bar{R},L}\big |_{x^{0}%
=0}\nonumber\\
&  =\int_{-\infty}^{+\infty}d_{1}^{n}p\int_{-\infty}^{+\infty}d_{1}^{n}%
x\exp(\text{i}^{-1}p_{i}|x^{j})_{\bar{R},L}\big |_{x^{0}=0},\\[0.1in]
\text{vol}_{2}  &  \equiv\int_{-\infty}^{+\infty}d_{2}^{n}x\int_{-\infty
}^{+\infty}d_{2}^{n}p\exp(x^{i}|\text{i}^{-1}p_{j})_{R,\bar{L}}\big |_{x^{0}%
=0}\nonumber\\
&  =\int\nolimits_{-\infty}^{+\infty}d_{2}^{n}p\int_{-\infty}^{+\infty}%
d_{2}^{n}x\exp(\text{i}^{-1}p_{i}|x^{j})_{R,\bar{L}}\big |_{x^{0}=0},
\end{align}
the orthonormality relations become%
\begin{align}
&  \big \langle(\bar{u}_{\bar{R},L})_{p,m}(x^{A},\pm t),(u_{\bar{R}%
,L})_{\ominus_{L}\tilde{p},m}(x^{B},\mp q^{-\zeta}t)\big \rangle_{1,x}%
^{\prime}=\nonumber\\
&  \hspace{0.4in}=\,\big \langle(u_{\bar{R},L})_{p,m}(x^{A},\pm t),(\bar
{u}_{\bar{R},L})_{\ominus_{\bar{R}}\tilde{p},m}(x^{B},\mp q^{-\zeta
}t)\big \rangle_{1,x}\nonumber\\
&  \hspace{0.4in}=\,\int_{-\infty}^{+\infty}d_{1}^{n}x\text{\thinspace}%
(\bar{u}_{\bar{R},L})_{p,m}(x^{A},\pm t)\overset{x|\tilde{p}}{\odot}%
_{\hspace{-0.01in}R}(\bar{u}_{\bar{R},L})_{\ominus_{\bar{R}}\tilde{p},m}%
(x^{B},\mp q^{-\zeta}t)\nonumber\\
&  \hspace{0.4in}=\,(\text{vol}_{1})^{-1}\delta_{1}^{n}(p_{C}\oplus_{\bar{R}%
}(\ominus_{\bar{R}}\,\tilde{p}_{D})),\label{OrthRel1}\\[0.1in]
&  \big \langle(\bar{u}_{R,\bar{L}})_{p,m}(x^{A},\pm t),(u_{R,\bar{L}%
})_{\ominus_{\bar{L}}\tilde{p},m}(x^{B},\mp q^{\zeta}t)\big \rangle_{2,x}%
^{\prime}=\nonumber\\
&  \hspace{0.4in}=\,\big \langle(u_{R,\bar{L}})_{p,m}(x^{A},\pm t),(\bar
{u}_{R,\bar{L}})_{\ominus_{R}\tilde{p},m}(x^{B},\mp q^{\zeta}%
t)\big \rangle_{2,x}\nonumber\\
&  \hspace{0.4in}=\,\int_{-\infty}^{+\infty}d_{2}^{n}x\text{\thinspace}%
(\bar{u}_{R,\bar{L}})_{p,m}(x^{A},\pm t)\overset{x|\tilde{p}}{\odot}%
_{\hspace{-0.01in}\bar{R}}(\bar{u}_{R,\bar{L}})_{\ominus_{R}\tilde{p},m}%
(x^{B},\mp q^{\zeta}t)\nonumber\\
&  \hspace{0.4in}=\,(\text{vol}_{2})^{-1}\delta_{2}^{n}(p_{C}\oplus
_{R}(\ominus_{R}\,\tilde{p}_{D})), \label{OrthRel2}%
\end{align}
and%
\begin{align}
&  \big \langle(u_{\bar{R},L})_{\ominus_{L}\tilde{p},m}(x^{A},\pm q^{-\zeta
}t),(\bar{u}_{\bar{R},L})_{p,m}(x^{B},\mp t)\big \rangle_{1,x}^{\prime
}=\nonumber\\
&  \hspace{0.4in}=\,\big \langle(\bar{u}_{\bar{R},L})_{\ominus_{\bar{R}}%
\tilde{p},m}(x^{A},\pm q^{-\zeta}t),(u_{\bar{R},L})_{p,m}(x^{B},\mp
t)\big \rangle_{1,x}\nonumber\\
&  \hspace{0.4in}=\,\int_{-\infty}^{+\infty}d_{1}^{n}p\text{\thinspace
}(u_{\bar{R},L})_{\ominus_{L}\tilde{p},m}(x^{A},\pm q^{-\zeta}t)\overset
{\tilde{p}|x}{\odot}_{\hspace{-0.01in}\bar{L}}(u_{p})_{\bar{R},L}(x^{B},\mp
t)\nonumber\\
&  \hspace{0.4in}=\,(\text{vol}_{1})^{-1}\delta_{1}^{n}((\ominus_{L}%
\,\tilde{p}_{C})\oplus_{L}p_{D}),\\[0.1in]
&  \big \langle(u_{R,\bar{L}})_{\ominus_{\bar{L}}\tilde{p},m}(x^{A},\pm
q^{\zeta}t),(\bar{u}_{R,\bar{L}})_{p,m}(x^{B},\mp t)\big \rangle_{2,x}%
^{\prime}=\nonumber\\
&  \hspace{0.4in}=\,\big \langle(\bar{u}_{R,\bar{L}})_{\ominus_{R}\tilde{p}%
,m}(x^{A},\pm q^{\zeta}t),(u_{R,\bar{L}})_{p,m}(x^{B},\mp t)\big \rangle_{2,x}%
\nonumber\\
&  \hspace{0.4in}=\,\int_{-\infty}^{+\infty}d_{2}^{n}p\text{\thinspace
}(u_{R,\bar{L}})_{\ominus_{\bar{L}}\tilde{p},m}(x^{A},\pm q^{\zeta}%
t)\overset{\tilde{p}|x}{\odot}_{\hspace{-0.01in}L}(u_{p})_{R,\bar{L}}%
(x^{B},\mp t)\nonumber\\
&  \hspace{0.4in}=\,(\text{vol}_{2})^{-1}\delta_{2}^{n}((\ominus_{\bar{L}%
}\,\tilde{p}_{C})\oplus_{\bar{L}}p_{D}). \label{OrthRel4}%
\end{align}

Clearly, the third equality in each of the above equations gives the explicit
form of the sesquilinear form on position space. The symbol on top of the
braided product indicates the tensor factors being involved in the braiding.
Expressions for calculating q-integrals over the whole space were given in
part I of our paper [cf. Sec.\thinspace5 of part I]. Notice that the
q-deformed delta functions are defined by \cite{Qkin1, KM94}
\begin{align}
\delta_{1}^{n}(p_{A}) &  \equiv\int_{-\infty}^{+\infty}d_{1}^{n}x\exp
(\text{i}^{-1}p_{k}|x^{j})_{\bar{R},L}\big |_{x^{0}=0},\nonumber\\
\delta_{2}^{n}(p_{A}) &  \equiv\int_{-\infty}^{+\infty}d_{2}^{n}%
x\,\exp(\text{i}^{-1}p_{k}|x^{j})_{R,\bar{L}}\big |_{x^{0}=0}.\label{DefDelt2}%
\end{align}
We recommend Refs. \cite{qAn, Qkin1, Qkin2} if the reader wants to have some
more information about our formalism.

Now, we come to the completeness relations for momentum eigenfunctions of
equal\ time. They take the form%
\begin{align}
&  \big \langle(u_{\bar{R},L})_{p,m}(x^{A},\pm t),(\bar{u}_{\bar{R}%
,L})_{\ominus_{\bar{R}}p,m}(y^{B},\mp q^{-\zeta}t)\big \rangle_{1,p}^{\prime
}=\nonumber\\
&  \hspace{0.4in}=\,\big \langle(\bar{u}_{\bar{R},L})_{p,m}(x^{A},\pm
t),(u_{\bar{R},L})_{\ominus_{L}p,m}(y^{B},\mp q^{-\zeta}t)\big \rangle_{1,p}%
\nonumber\\
&  \hspace{0.4in}=\,\int_{-\infty}^{+\infty}d_{1}^{n}p\text{\thinspace
}(u_{\bar{R},L})_{p,m}(x^{A},\pm t)\overset{p|y}{\odot}_{\hspace{-0.01in}%
R}(u_{\bar{R},L})_{\ominus_{L}p,m}(y^{B},\mp q^{-\zeta}t)\nonumber\\
&  \hspace{0.4in}=\,(\text{vol}_{1})^{-1}\delta_{1}^{n}(x^{A}\oplus_{\bar{R}%
}(\ominus_{\bar{R}}\,y^{B})),\label{ComRelP1}\\[0.1in]
&  \big \langle(u_{R,\bar{L}})_{p,m}(x^{A},\pm t),(\bar{u}_{R,\bar{L}%
})_{\ominus_{R}p,m}(y^{B},\mp q^{\zeta}t)\big \rangle_{2,p}^{\prime
}=\nonumber\\
&  \hspace{0.4in}=\,\big \langle(\bar{u}_{R,\bar{L}})_{p,m}(x^{A},\pm
t),(u_{R,\bar{L}})_{\ominus_{\bar{L}}p,m}(y^{B},\mp q^{\zeta}%
t)\big \rangle_{2,p}\nonumber\\
&  \hspace{0.4in}=\,\int_{-\infty}^{+\infty}d_{2}^{n}p\text{\thinspace
}(u_{R,\bar{L}})_{p,m}(x^{A},\pm t)\overset{p|y}{\odot}_{\hspace{-0.01in}%
\bar{R}}(u_{R,\bar{L}})_{\ominus_{\bar{L}}p,m}(y^{B},\mp q^{\zeta
}t)\nonumber\\
&  \hspace{0.4in}=\,(\text{vol}_{2})^{-1}\delta_{2}^{n}(x^{A}\oplus
_{R}(\ominus_{R}\,y^{B})), \label{ComRelP2}%
\end{align}
and%
\begin{align}
&  \big \langle(\bar{u}_{\bar{R},L})_{\ominus_{\bar{R}}p,m}(y^{B},\pm
q^{-\zeta}t),(u_{\bar{R},L})_{p,m}(x^{A},\mp t)\big \rangle_{2,p}^{\prime
}=\nonumber\\
&  \hspace{0.4in}=\,\big \langle(u_{\bar{R},L})_{\ominus_{L}p,m}(y^{B},\pm
q^{-\zeta}t),(\bar{u}_{\bar{R},L})_{p,m}(x^{A},\mp t)\big \rangle_{1,p}%
\nonumber\\
&  \hspace{0.4in}=\,\int_{-\infty}^{+\infty}d_{1}^{n}p\text{\thinspace}%
(\bar{u}_{\bar{R},L})_{\ominus_{\bar{R}}p,m}(y^{B},\pm q^{-\zeta}%
t)\overset{y|p}{\odot}_{\hspace{-0.01in}R}(\bar{u}_{\bar{R},L})_{p,m}%
(x^{A},\mp t)\nonumber\\
&  \hspace{0.4in}=\,(\text{vol}_{1})^{-1}\delta_{1}^{n}((\ominus_{L}%
\,y^{B})\oplus_{L}x^{A}),\\[0.1in]
&  \big \langle(\bar{u}_{\ominus_{R}p,m})_{R,\bar{L}}(y^{B},\pm q^{\zeta
}t),(u_{R,\bar{L}})_{p,m}(x^{A},\mp t)\big \rangle_{2,p}^{\prime}=\nonumber\\
&  \hspace{0.4in}=\,\big \langle(u_{R,\bar{L}})_{\ominus_{\bar{L}}p,m}%
(y^{B},\pm q^{\zeta}t),(\bar{u}_{R,\bar{L}})_{p,m}(x^{A},\mp
t)\big \rangle_{2,p}\nonumber\\
&  \hspace{0.4in}=\,\int_{-\infty}^{+\infty}d_{2}^{n}p\text{\thinspace}%
(\bar{u}_{R,\bar{L}})_{\ominus_{R}p,m}(y^{B},\pm q^{\zeta}t)\overset
{y|p}{\odot}_{\hspace{-0.01in}\bar{R}}(\bar{u}_{R,\bar{L}})_{p,m}(x^{A},\mp
t)\nonumber\\
&  \hspace{0.4in}=\,(\text{vol}_{2})^{-1}\delta_{2}^{n}((\ominus_{\bar{L}%
}\,y^{B})\oplus_{\bar{L}}x^{A}).
\end{align}

To check completeness and orthonormality of momentum eigenfunctions we first
recall that in Ref. \cite{Qkin2}\textbf{ }it was already shown that the above
relations are valid for $t=0.$ However, the time-dependence of momentum
eigenfunctions results from phase factors. Due to their algebraic properties
these phase factors can be brought together in such a way that they cancel
each other out. The last assertion can easily\ be checked by direct inspection
of the relations in (\ref{ExpTimDep1}), (\ref{ExpTimDep2}), (\ref{ExpTimDep3}%
), and (\ref{ExpTimDep4}).

We would like to illustrate these reasonings by the following calculation:%
\begin{align}
&  \int_{-\infty}^{+\infty}d_{1}^{n}p\text{\thinspace}(\bar{u}_{\bar{R}%
,L})_{\ominus_{\bar{R}}p,m}(y^{B},q^{-\zeta}t)\overset{y|p}{\odot}%
_{\hspace{-0.01in}R}(\bar{u}_{\bar{R},L})_{p,m}(x^{A},-t)=\nonumber\\
&  \qquad=\int_{-\infty}^{+\infty}d_{1}^{n}p\text{\thinspace}\big (\exp
(\text{i}(2m)^{-1}p^{2}t)_{\bar{R},L}\overset{p}{\circledast}(\bar{u}_{\bar
{R},L})_{\ominus_{\bar{R}}p,m}(y^{B},0)\big )\nonumber\\
&  \qquad\qquad\qquad\overset{y|p}{\odot}_{\hspace{-0.01in}R}\big (\exp
(-\text{i}(2m)^{-1}p^{2}t)_{\bar{R},L}\overset{p}{\circledast}(\bar{u}%
_{\bar{R},L})_{p,m}(x^{A},0)\big )\nonumber\\
&  \qquad=\int_{-\infty}^{+\infty}d_{1}^{n}p\text{\thinspace}(\bar{u}_{\bar
{R},L})_{\ominus_{\bar{R}}p,m}(y^{B},0)\nonumber\\
&  \qquad\qquad\qquad\overset{y|p}{\odot}_{\hspace{-0.01in}R}\big (\exp
(\text{i}(2m)^{-1}p^{2}t)_{\bar{R},L}\overset{p}{\circledast}\exp
(-\text{i}(2m)^{-1}p^{2}t)_{\bar{R},L}\nonumber\\
&  \qquad\qquad\qquad\overset{p}{\circledast}(\bar{u}_{\bar{R},L})_{p,m}%
(x^{A},0)\big )\nonumber\\
&  \qquad=\,\int_{-\infty}^{+\infty}d_{1}^{n}p\text{\thinspace}(\bar{u}%
_{\bar{R},L})_{\ominus_{\bar{R}}p,m}(y^{B},0)\overset{y|p}{\odot}%
_{\hspace{-0.01in}R}(\bar{u}_{\bar{R},L})_{p,m}(x^{A},0)\nonumber\\
&  \qquad=\,(\text{vol}_{1})^{-1}\delta_{1}^{n}((\ominus_{L}\,y^{B})\oplus
_{L}x^{A}).
\end{align}
For the first step we make use of the relations in (\ref{ExpTimDep2}) and
(\ref{ExpTimDep4}). Then we rearrange terms by taking into account that the
time-dependent phase factors commute with all other factors. For the third
step we have to realize that the two phase factors are inverse to each other.
The last equality is the completeness relation for time-independent momentum
eigenfunctions as it was derived in Ref. \cite{Qkin1}.

In classical quantum mechanics momentum eigenfunctions are not elements of a
Hilbert space, since they are not square-integrable functions. Instead,
elements of a Hilbert space are obtained by linear superposition of stationary
momentum eigenfunctions. In other words, physical states are represented by
so-called wave packets. From the results in Ref. \cite{Qkin2}\textbf{ }we can
read off the explicit form of these wave packets. Again, we can directly apply
these reasonings, so we get
\begin{align}
(\phi_{1})_{m}^{\prime}(x^{i})  &  =\frac{\kappa^{n}}{(\text{vol}_{1})^{1/2}%
}\int_{-\infty}^{+\infty}d_{1}^{n}p\,(c_{1})_{\kappa p}^{\prime}\overset
{p|x}{\odot}_{\hspace{-0.01in}R}(u_{\bar{R},L})_{\ominus_{L}p,m}%
(x^{i})\nonumber\\
&  =\frac{\kappa^{n}}{(\text{vol}_{1})^{1/2}}\big \langle(c_{1})_{\kappa
p}^{\prime},(\bar{u}_{\bar{R},L})_{\ominus_{\bar{R}}p,m}(x^{i}%
)\big \rangle_{1,p}^{\prime},\label{EntImp1}\\[0.1in]
(\phi_{2})_{m}(x^{i})  &  =\frac{\kappa^{-n}}{(\text{vol}_{2})^{1/2}}%
\int_{-\infty}^{+\infty}d_{2}^{n}p\,(\bar{u}_{R,\bar{L}})_{\ominus_{R}%
p,m}(x^{i})\overset{x|p}{\odot}_{\hspace{-0.01in}L}(c_{2})_{\kappa^{-1}%
p}\nonumber\\
&  =\frac{\kappa^{-n}}{(\text{vol}_{2})^{1/2}}\big \langle(u_{R,\bar{L}%
})_{\ominus_{\bar{L}}p,m}(x^{i}),(c_{2})_{\kappa^{-1}p}\big \rangle_{2,p},
\end{align}
and%
\begin{align}
(\phi_{1}^{\ast})_{m}(x^{i})  &  =(\text{vol}_{1})^{1/2}\int_{-\infty
}^{+\infty}d_{1}^{n}p\,(u_{\bar{R},L})_{p,m}(x^{i})\overset{p}{\circledast
}(c_{1}^{\ast})_{\kappa^{-1}p}\nonumber\\
&  =(\text{vol}_{1})^{1/2}\big \langle(\bar{u}_{\bar{R},L})_{p,m}%
(x^{i}),(c_{1}^{\ast})_{\kappa^{-1}p}\big \rangle_{1,p},\\[0.1in]
(\phi_{2}^{\ast})_{m}^{\prime}(x^{i})  &  =(\text{vol}_{2})^{1/2}\int
_{-\infty}^{+\infty}d_{2}^{n}p\,(c_{2}^{\ast})_{\kappa p}^{\prime}\overset
{p}{\circledast}(\bar{u}_{R,\bar{L}})_{p,m}(x^{i})\nonumber\\
&  =(\text{vol}_{2})^{1/2}\big \langle(c_{2}^{\ast})_{\kappa p}^{\prime
},(u_{R,\bar{L}})_{p,m}(x^{i})\big \rangle_{2,p}^{\prime}, \label{EntIm2}%
\end{align}
where the constant $\kappa$ takes on as values

\begin{enumerate}
\item[(i)] (braided line) $\kappa=q,$

\item[(ii)] (q-deformed Euclidean space in three dimensions) $\kappa=q^{6}.$
\end{enumerate}

Applying the substitutions
\begin{equation}
L\leftrightarrow\bar{L},\quad R\leftrightarrow\bar{R},\quad\kappa
\leftrightarrow\kappa^{-1},\quad1\text{ (as label)}\leftrightarrow2\text{ (as
label)}, \label{TransRule3}%
\end{equation}
to the formulae in (\ref{EntImp1})-(\ref{EntIm2}) yields further expressions
for wave packets. The existence of such crossing-symmetries is a typical
feature of q-deformation (see for example Ref. \cite{qAn}).

The wave-packets in (\ref{EntImp1})-(\ref{EntIm2}) give solutions to the
free-particle Schr\"{o}dinger equations in (\ref{FreParSch1}) and
(\ref{FreParSch2}). Using the equations in (\ref{FreSchGlImp1}) and
(\ref{FreSchGlImp4}) one readily checks that%
\begin{align}
\text{i}\partial_{0}\overset{t}{\triangleright}(\phi_{1})_{m}^{\prime}(x^{i})
&  =H_{0}\overset{x}{\triangleright}(\phi_{1})_{m}^{\prime}(x^{i}),\nonumber\\
\text{i}\partial_{0}\overset{t}{\triangleright}(\phi_{1}^{\ast})_{m}(x^{i})
&  =H_{0}\overset{x}{\triangleright}(\phi_{1}^{\ast})_{m}(x^{i}),\\[0.1in]
\text{i}\hat{\partial}_{0}\,\overset{t}{\bar{\triangleright}}\,(\phi_{2}%
)_{m}^{\prime}(x^{i})  &  =H_{0}\,\overset{x}{\bar{\triangleright}}\,(\phi
_{2})_{m}^{\prime}(x^{i}),\nonumber\\
\text{i}\hat{\partial}_{0}\,\overset{t}{\bar{\triangleright}}\,(\phi_{2}%
^{\ast})_{m}(x^{i})  &  =H_{0}\,\overset{x}{\bar{\triangleright}}\,(\phi
_{2}^{\ast})_{m}(x^{i}),
\end{align}
and
\begin{align}
(\phi_{2})_{m}(x^{i})\overset{t}{\triangleleft}(\text{i}\hat{\partial}_{0})
&  =(\phi_{2})_{m}(x^{i})\overset{x}{\triangleleft}H_{0},\nonumber\\
(\phi_{2}^{\ast})_{m}^{\prime}(x^{i})\overset{t}{\triangleleft}(\text{i}%
\hat{\partial}_{0})  &  =(\phi_{2}^{\ast})_{m}^{\prime}(x^{i})\overset
{x}{\triangleleft}H_{0},\\[0.1in]
(\phi_{1})_{m}(x^{i})\,\overset{t}{\bar{\triangleleft}}\,(\text{i}\partial
_{0})  &  =(\phi_{1})_{m}(x^{i})\,\overset{x}{\bar{\triangleleft}}%
\,H_{0},\nonumber\\
(\phi_{1}^{\ast})_{m}^{\prime}(x^{i})\,\overset{t}{\bar{\triangleleft}%
}\,(\text{i}\partial_{0})  &  =(\phi_{1}^{\ast})_{m}^{\prime}(x^{i}%
)\,\overset{x}{\bar{\triangleleft}}\,H_{0}.
\end{align}

At this place it should be mentioned that $(\phi_{i})_{m}^{\prime}$ and
$(\phi_{i}^{\ast})_{m}$ traverse forward in time, while\textbf{ }$(\phi
_{i})_{m}$ and $(\phi_{i}^{\ast})_{m}^{\prime}$ move oppositely with time.
Concretely, the time evolution of these wave packets is determined by%
\begin{align}
(\phi_{1})_{m}^{\prime}(x^{i})  &  =\exp(-t\otimes\text{i}H_{0})\overset
{H_{0}|x}{\triangleright}(\phi_{1})_{m}^{\prime}(x^{A},t=0),\nonumber\\
(\phi_{1}^{\ast})_{m}(x^{i})  &  =\exp(-t\otimes\text{i}H_{0})\overset
{H_{0}|x}{\triangleright}(\phi_{1}^{\ast})_{m}(x^{A},t=0), \label{TimEntEbe1}%
\\[0.1in]
(\phi_{2})_{m}^{\prime}(x^{i})  &  =\exp(-t\otimes\text{i}H_{0})\,\overset
{H_{0}|x}{\bar{\triangleright}}\,(\phi_{2})_{m}^{\prime}(x^{A}%
,t=0),\nonumber\\
(\phi_{2}^{\ast})_{m}(x^{i})  &  =\exp(-t\otimes\text{i}H_{0})\,\overset
{H_{0}|x}{\bar{\triangleright}}\,(\phi_{2}^{\ast})_{m}(x^{A},t=0),
\label{TimEntEbe2}%
\end{align}
and%
\begin{align}
(\phi_{2})_{m}(x^{i})  &  =(\phi_{2})_{m}(x^{A},t=0)\overset{x|H_{0}%
}{\triangleleft}\exp(\text{i}H_{0}\otimes t),\nonumber\\
(\phi_{2}^{\ast})_{m}^{\prime}(x^{i})  &  =(\phi_{2}^{\ast})_{m}^{\prime
}(x^{A},t=0)\overset{x|H_{0}}{\triangleleft}\exp(\text{i}H_{0}\otimes
t),\label{TimEntEbe3}\\[0.1in]
(\phi_{1})_{m}(x^{i})  &  =(\phi_{1})_{m}(x^{A},t=0)\,\overset{x|H_{0}}%
{\bar{\triangleleft}}\,\exp(\text{i}H_{0}\otimes t),\nonumber\\
(\phi_{1}^{\ast})_{m}^{\prime}(x^{i})  &  =(\phi_{1}^{\ast})_{m}^{\prime
}(x^{A},t=0)\,\overset{x|H_{0}}{\bar{\triangleleft}}\,\exp(\text{i}%
H_{0}\otimes t). \label{TimEntEbe4}%
\end{align}

A short glance at the expansions in (\ref{EntImp1})-(\ref{EntIm2}) shows us
that they are written in terms of time-dependent momentum eigenfunctions. In
the Schr\"{o}dinger picture, however, operators are assumed to be independent
from time and the same should hold for their eigenfunctions. Thus, to give the
expansion coefficients a physical meaning it is convenient to reformulate the
expansions in (\ref{EntImp1})-(\ref{EntIm2}) in terms of time-independent
momentum states. In doing so, we find
\begin{align}
(\phi_{1})_{m}^{\prime}(x^{i})  &  =\frac{\kappa^{n}}{(\text{vol}_{1})^{1/2}%
}\int_{-\infty}^{+\infty}d_{1}^{n}p\,(c_{1})_{\kappa p}^{\prime}(\kappa
^{-2}t)\overset{p|x}{\odot}_{\hspace{-0.01in}R}(u_{\bar{R},L})_{\ominus
_{L}p,m}(x^{A},t=0)\nonumber\\
&  =\kappa^{n}(\text{vol}_{1})^{-1/2}\big \langle(c_{1})_{\kappa p}^{\prime
}(\kappa^{-2}t),(\bar{u}_{\bar{R},L})_{\ominus_{\bar{R}}p,m}(x^{A}%
,t=0)\big \rangle_{1,p}^{\prime},\label{Exp1}\\[0.1in]
(\phi_{2})_{m}(x^{i})  &  =\frac{\kappa^{-n}}{(\text{vol}_{2})^{1/2}}%
\int_{-\infty}^{+\infty}d_{2}^{n}p\,(\bar{u}_{R,\bar{L}})_{\ominus_{R}%
p,m}(x^{A},t=0)\overset{x|p}{\odot}_{\hspace{-0.01in}L}(c_{2})_{\kappa^{-1}%
p}(\kappa^{2}t)\nonumber\\
&  =\kappa^{-n}(\text{vol}_{2})^{-1/2}\big \langle(u_{R,\bar{L}}%
)_{\ominus_{\bar{L}}p,m}(x^{A},t=0),(c_{2})_{\kappa^{-1}p}(\kappa
^{2}t)\big \rangle_{2,p},
\end{align}
and%
\begin{align}
(\phi_{1}^{\ast})_{m}(x^{i})  &  =(\text{vol}_{1})^{1/2}\int_{-\infty
}^{+\infty}d_{1}^{n}p\,(u_{\bar{R},L})_{p,m}(x^{A},t=0)\overset{p}%
{\circledast}(c_{1}^{\ast})_{\kappa^{-1}p}(\kappa^{2}t)\nonumber\\
&  =(\text{vol}_{1})^{1/2}\big \langle(\bar{u}_{\bar{R},L})_{p,m}%
(x^{A},t=0),(c_{1}^{\ast})_{\kappa^{-1}p}(\kappa^{2}t)\big \rangle_{1,p}%
,\\[0.1in]
(\phi_{2}^{\ast})_{m}^{\prime}(x^{i})  &  =(\text{vol}_{2})^{1/2}\int
_{-\infty}^{+\infty}d_{2}^{n}p\,(c_{2}^{\ast})_{\kappa p}^{\prime}(\kappa
^{-2}t)\overset{p}{\circledast}(\bar{u}_{R,\bar{L}})_{p,m}(x^{A}%
,t=0)\nonumber\\
&  =(\text{vol}_{2})^{1/2}\big \langle(c_{2}^{\ast})_{\kappa p}^{\prime
}(\kappa^{-2}t),(u_{R,\bar{L}})_{p,m}(x^{i},t=0)\big \rangle_{2,p}^{\prime},
\label{Exp4}%
\end{align}
where the time-dependent expansion coefficients are now given by
\begin{align}
(c_{1})_{p}^{\prime}(t)  &  =(c_{1})_{p}^{\prime}\overset{p}{\circledast}%
\exp(-\text{i}q^{\zeta}tp^{2}(2m)^{-1})_{\bar{R},L},\nonumber\\
(c_{2}^{\ast})_{p}^{\prime}(t)  &  =(c_{2}^{\ast})_{p}\overset{p}{\circledast
}\exp(\text{i}(2m)^{-1}p^{2}t)_{R,\bar{L}},\label{CtCexp1}\\[0.1in]
(c_{2})_{p}(t)  &  =\exp(\text{i}q^{-\zeta}(2m)^{-1}p^{2}t)_{R,\bar{L}%
}\overset{p}{\circledast}(c_{2})_{p},\nonumber\\
(c_{1}^{\ast})_{p}(t)  &  =\exp(-\text{i}tp^{2}(2m)^{-1})_{\bar{R},L}%
\overset{p}{\circledast}(c_{1}^{\ast})_{p}. \label{CtCexp2}%
\end{align}
Notice that the last equalities are a direct consequence of the identities in
(\ref{ExpTimDep1}) and (\ref{ExpTimDep2}), if we take into account the trivial
braiding of the time-dependent phase factors.

Eqs. (\ref{Exp1})-(\ref{Exp4}) are nothing other than Fourier expansions of
the solutions to the free-particle Schr\"{o}dinger equations in
(\ref{FreParSch1}) and (\ref{FreParSch2}). In this respect, the coefficients
$(c_{i})_{p}(t),$ $(c_{i})_{p}^{\prime}(t),$ $(c_{i}^{\ast})_{p}(t),$ and
$(c_{i}^{\ast})_{p}^{\prime}(t)$ can be viewed as probability amplitudes for
observing a particle of definite momentum at time $t$ (see also the discussion
in Ref. \cite{Qkin2}). The reasonings about inverse Fourier transformations in
Ref. \cite{Qkin1} showed us how to calculate these probability amplitudes from
the corresponding wave packets. In this manner, we have
\begin{align}
(c_{1})_{p}^{\prime}(t)  &  =(\text{vol}_{1})^{1/2}\int\nolimits_{-\infty
}^{+\infty}d_{1}^{n}x\,(\phi_{1})_{m}^{\prime}(x^{i})\overset{x}{\circledast
}(u_{\bar{R},L})_{p,m}(x^{A},t=0)\nonumber\\
&  =(\text{vol}_{1})^{1/2}\big \langle(\phi_{1})_{m}^{\prime}(x^{i}),(\bar
{u}_{\bar{R},L})_{p,m}(x^{A},t=0)\big \rangle_{1,x}^{\prime}, \label{cLpt}%
\\[0.1in]
(c_{2})_{p}(t)  &  =(\text{vol}_{2})^{1/2}\int\nolimits_{-\infty}^{+\infty
}d_{2}^{n}x\,(\bar{u}_{R,\bar{L}})_{p,m}(x^{A},t=0)\overset{x}{\circledast
}(\phi_{2})_{m}(x^{i})\nonumber\\
&  =(\text{vol}_{2})^{1/2}\big \langle(u_{R,\bar{L}})_{p,m}(x^{A}%
,t=0),(\phi_{2})_{m}(x^{i})\big \rangle_{2,x},
\end{align}
and%
\begin{align}
(c_{1}^{\ast})_{p}(t)  &  =\frac{1}{(\text{vol}_{1})^{1/2}}\int
\nolimits_{-\infty}^{+\infty}d_{1}^{n}x\,(u_{\bar{R},L})_{\ominus_{\bar{R}%
}p,m}(x^{A},t=0)\overset{p|x}{\odot}_{\hspace{-0.01in}\bar{L}}(\phi_{1}^{\ast
})_{m}(x^{i})\nonumber\\
&  =\frac{1}{(\text{vol}_{1})^{1/2}}\big \langle(\bar{u}_{\bar{R},L}%
)_{\ominus_{L}p,m}(x^{A},t=0),(\phi_{1}^{\ast})_{m}(x^{i})\big \rangle_{1,x}%
,\\[0.1in]
(c_{2}^{\ast})_{p}^{\prime}(t)  &  =\frac{1}{(\text{vol}_{2})^{1/2}}%
\int\nolimits_{-\infty}^{+\infty}d_{2}^{n}x\,(\phi_{2}^{\ast})_{m}^{\prime
}(x^{i})\overset{x|p}{\odot}_{\hspace{-0.01in}\bar{R}}(\bar{u}_{R,\bar{L}%
})_{\ominus_{\bar{L}}p,m}(x^{A},t=0)\nonumber\\
&  =\frac{1}{(\text{vol}_{2})^{1/2}}\big \langle(\phi_{2}^{\ast})_{m}^{\prime
}(x^{i}),(u_{R,\bar{L}})_{\ominus_{R}p,m}(x^{A},t=0)\big \rangle_{2,x}%
^{\prime}.
\end{align}

As already mentioned, the above formulae for computing the expansion
coefficients refer to the Schr\"{o}dinger picture. This can be seen from the
observation that the momentum eigenfunctions are fixed in time. In the
Heisenberg picture, however, observables together with their eigenfunctions
vary with time. Fortunately, the transition to the Heisenberg picture\textbf{
}can easily be achieved by exploiting the conjugation properties of time
evolution operators.

We would like to illustrate this assertion by the following calculation:%
\begin{align}
(c_{1}^{\ast})_{p}(t)  &  =(\text{vol}_{1})^{-1/2}\big \langle(\bar{u}%
_{\bar{R},L})_{\ominus_{L}p,m}(x^{A},t=0),(\phi_{1}^{\ast})_{m}(x^{i}%
)\big \rangle_{1,x}\nonumber\\
&  =(\text{vol}_{1})^{-1/2}\big \langle(\bar{u}_{\bar{R},L})_{\ominus_{L}%
p,m}(x^{A},0),\exp(-\text{i}tH_{0})\overset{H_{0}|x}{\triangleright}(\phi
_{1}^{\ast})_{m}(x^{A},0)\big \rangle_{1,x}\nonumber\\
&  =(\text{vol}_{1})^{-1/2}\big \langle\exp(\text{i}tH_{0})\,\overset{H_{0}%
|x}{\bar{\triangleright}}\,(\bar{u}_{\bar{R},L})_{\ominus_{L}p,m}%
(x^{A},0),(\phi_{1}^{\ast})_{m}(x^{A},0)\big \rangle_{1,x}\nonumber\\
&  =(\text{vol}_{1})^{-1/2}\big \langle(\bar{u}_{\bar{R},L})_{\ominus_{L}%
p,m}(x^{A},0)\,\overset{x|H_{0}}{\bar{\triangleleft}}\!\exp(\text{i}q^{-\zeta
}tH_{0}),(\phi_{1}^{\ast})_{m}(x^{A},0)\big \rangle_{1,x}\nonumber\\
&  =(\text{vol}_{1})^{-1/2}\big \langle(\bar{u}_{\bar{R},L})_{\ominus_{L}%
p,m}(x^{A},q^{-\zeta}t),(\phi_{1}^{\ast})_{m}(x^{A},0)\big \rangle_{1,x}.
\end{align}
The second equality in the above calculation holds due to the second relation
in (\ref{TimEntEbe1}). For the third equality we use fact that the adjoint of
the time evolution operator is given by its Hermitian conjugate. Then we
express the left action of the time evolution operator by a right one. The
last equality can be recognized as the first identity in (\ref{ConPlaWav}). In
very much the same way we get
\begin{equation}
(c_{2}^{\ast})_{p}^{\prime}(t)=(\text{vol}_{2})^{-1/2}\big \langle(\phi
_{2}^{\ast})_{m}^{\prime}(x^{A},0),(u_{R,\bar{L}})_{\ominus_{R}p,m}%
(x^{A},q^{\zeta}t)\big \rangle_{2,x},
\end{equation}
and%
\begin{align}
(c_{1})_{p}^{\prime}(t)  &  =(\text{vol}_{1})^{1/2}\big \langle(\phi_{1}%
)_{m}^{\prime}(x^{A},0),(\bar{u}_{\bar{R},L})_{p,m}(x^{A},q^{\zeta
}t)\big \rangle_{1,x}^{\prime},\nonumber\\
(c_{2})_{p}(t)  &  =(\text{vol}_{2})^{1/2}\big \langle(u_{R,\bar{L}}%
)_{p,m}(x^{A},q^{-\zeta}t),(\phi_{2})_{m}(x^{A},0)\big \rangle_{2,x}.
\end{align}

From the considerations so far we can see that for one and the same wave
function there exist different expansions in terms of momentum eigenfunctions.
This is a consequence of the fact that we can distinguish different
q-geometries. Perhaps, the reader may have noticed that we often restrict
attention to certain q-geometries, only. The reason for this lies in the fact
that we can make transitions between the expressions corresponding to
different geometries by means of the substitutions%
\begin{gather}
L\leftrightarrow\bar{L},\quad R\leftrightarrow\bar{R},\quad\kappa
\leftrightarrow\kappa^{-1},\quad q\leftrightarrow q^{-1},\nonumber\\
1\text{ (as label)}\leftrightarrow2\text{ (as label)},\quad\partial
\leftrightarrow\hat{\partial},\quad\triangleright\leftrightarrow
\bar{\triangleright},\quad\triangleleft\leftrightarrow\bar{\triangleleft}.
\label{SubGeo}%
\end{gather}
Thus, it is sufficient to treat some q-geometries explicitly, since the
expressions for the other ones are easily obtained via the substitutions in
(\ref{SubGeo}).

\subsection{Probability densities and expectation values}

The existence of different q-geometries enables us to write down different
versions of the normalization condition of wave functions. In Ref. \cite{qAn}
we discussed the normalization conditions for wave functions on position as
well as momentum space. Adapting these ideas for our results gives
\begin{align}
1  &  =\big \langle\phi,\phi\big \rangle_{1,x}=\,\frac{1}{2}\big \langle(\phi
_{1})_{m}(-q^{-\zeta}t),(\phi_{1}^{\ast})_{m}(t)\big \rangle_{1,x}\nonumber\\
&  \qquad\qquad\qquad+\frac{1}{2}\big \langle(\phi_{1}^{\ast})_{m}%
(t),(\phi_{1})_{m}(-q^{-\zeta}t)\big \rangle_{1,x}\nonumber\\
&  =\int_{-\infty}^{+\infty}d_{1}\,p\,\frac{1}{2}\big (\,\overline{(c_{1}%
)_{p}(-q^{-\zeta}t)}\overset{p}{\circledast}(c_{1}^{\ast})_{\kappa^{-1}%
p}(\kappa^{2}t)\nonumber\\
&  \qquad\qquad\qquad+\overline{(c_{1}^{\ast})_{\kappa^{-1}p}(\kappa^{2}%
t)}\overset{p}{\circledast}(c_{1})_{p}(-q^{-\zeta}t)\big )\nonumber\\
&  =\int_{-\infty}^{+\infty}d_{1}\,p\,\frac{1}{2}\big (\,\overline{(c_{1}%
)_{p}}\overset{p}{\circledast}(c_{1}^{\ast})_{\kappa^{-1}p}+\overline
{(c_{1}^{\ast})_{\kappa^{-1}p}}\overset{p}{\circledast}(c_{1})_{p}%
\big ),\label{NorCon1}\\[0.1in]
1  &  =\big \langle\phi,\phi\big \rangle_{2,x}=\frac{1}{2}\big \langle(\phi
_{2})_{m}(-q^{\zeta}t),(\phi_{2}^{\ast})_{m}(t)\big \rangle_{2,x}\nonumber\\
&  \qquad\qquad\qquad+\frac{1}{2}\big \langle(\phi_{2}^{\ast})_{m}%
(t),(\phi_{2})_{m}(-q^{\zeta}t)\big \rangle_{2,x}\nonumber\\
&  =\int_{-\infty}^{+\infty}d_{2}\,p\,\frac{1}{2}\big (\,\overline{(c_{2}%
)_{p}(-q^{\zeta}t)}\overset{p}{\circledast}(c_{2}^{\ast})_{\kappa p}%
(\kappa^{-2}t)\nonumber\\
&  \qquad\qquad\qquad+\overline{(c_{2}^{\ast})_{\kappa p}(\kappa^{-2}%
t)}\overset{p}{\circledast}(c_{2})_{p}(-q^{\zeta}t)\big )\nonumber\\
&  =\int_{-\infty}^{+\infty}d_{2}\,p\,\frac{1}{2}\big (\,\overline{(c_{2}%
)_{p}}\overset{p}{\circledast}(c_{2}^{\ast})_{\kappa p}+\overline{(c_{2}%
^{\ast})_{\kappa p}}\overset{p}{\circledast}(c_{2})_{p}\big ),
\end{align}
and%
\begin{align}
1  &  =\big \langle\phi,\phi\big \rangle_{1,x}^{\prime}=\frac{1}%
{2}\big \langle(\phi_{1})_{m}^{\prime}(q^{-\zeta}t),(\phi_{1}^{\ast}%
)_{m}^{\prime}(-t)\big \rangle_{1,x}^{\prime}\nonumber\\
&  \qquad\qquad\qquad+\frac{1}{2}\big \langle(\phi_{1}^{\ast})_{m}^{\prime
}(-t),(\phi_{1})_{m}^{\prime}(q^{-\zeta}t)\big \rangle_{1,x}^{\prime
}\nonumber\\
&  =\int_{-\infty}^{+\infty}d_{1}\,p\,\frac{1}{2}\big ((c_{1})_{p}^{\prime
}(q^{-\zeta}t)\overset{p}{\circledast}\overline{(c_{1}^{\ast})_{\kappa^{-1}%
p}^{\prime}(-\kappa^{2}t)}\nonumber\\
&  \qquad\qquad\qquad+(c_{1}^{\ast})_{\kappa^{-1}p}^{\prime}(-\kappa
^{2}t)\overset{p}{\circledast}\overline{(c_{1})_{p}^{\prime}(q^{-\zeta}%
t)}\,\big )\nonumber\\
&  =\int_{-\infty}^{+\infty}d_{1}\,p\,\frac{1}{2}\big ((c_{1})_{p}^{\prime
}\overset{p}{\circledast}\overline{(c_{1}^{\ast})_{\kappa^{-1}p}^{\prime}%
}+(c_{1}^{\ast})_{\kappa^{-1}p}^{\prime}\overset{p}{\circledast}%
\overline{(c_{1})_{p}^{\prime}}\,\big ),\\[0.1in]
1  &  =\big \langle\phi,\phi\big \rangle_{2,x}^{\prime}=\frac{1}%
{2}\big \langle(\phi_{2})_{m}^{\prime}(q^{\zeta}t),(\phi_{2}^{\ast}%
)_{m}^{\prime}(-t)\big \rangle_{2,x}^{\prime}\nonumber\\
&  \qquad\qquad\qquad+\frac{1}{2}\big \langle(\phi_{2}^{\ast})_{m}^{\prime
}(-t),(\phi_{2})_{m}^{\prime}(q^{\zeta}t)\big \rangle_{2,x}^{\prime
}\nonumber\\
&  =\int_{-\infty}^{+\infty}d_{2}\,p\,\frac{1}{2}\big ((c_{2})_{p}^{\prime
}(q^{\zeta}t)\overset{p}{\circledast}\overline{(c_{2}^{\ast})_{\kappa
p}^{\prime}(-\kappa^{-2}t)}\nonumber\\
&  \qquad\qquad\qquad+(c_{2}^{\ast})_{\kappa p}^{\prime}(-\kappa
^{-2}t)\overset{p}{\circledast}\overline{(c_{2})_{p}^{\prime}(q^{\zeta}%
t)}\,\big )\nonumber\\
&  =\int_{-\infty}^{+\infty}d_{2}\,p\,\frac{1}{2}\big ((c_{2})_{p}^{\prime
}\overset{p}{\circledast}\overline{(c_{2}^{\ast})_{\kappa p}^{^{\prime}}%
}+(c_{2}^{\ast})_{\kappa p}^{\prime}\overset{p}{\circledast}\overline
{(c_{2})_{p}^{\prime}}\,\big ).
\end{align}

Notice that the minus signs before the time variables are a consequence of the
fact that the conjugate solutions to the free Schr\"{o}dinger equations move
backwards in time. The last equality in each of the above equations follows
from the identities in (\ref{CtCexp1}) and (\ref{CtCexp2}). It tells us that
the normalization of a free-particle wave function does not change in time. To
get this result we have to choose the correct expansion for each argument of
our sesquilinear forms. Furthermore, the time coordinate of one wave function
as argument of a sesquilinear form has to be rescaled by a suitable factor.

Next, we come to the expectation values of momentum for a free particle. Using
the results of Ref. \cite{qAn} once more, we get%
\begin{align}
&  \big \langle\frac{1}{2}(P_{A}+\overline{P_{A}}\,)_{\phi}\big \rangle_{1,x}%
=\big \langle\phi,\frac{1}{2}(\text{i}\partial_{A}\overset{x}{\triangleright
}\phi+\overline{\text{i}\partial_{A}}\,\overset{x}{\bar{\triangleright}}%
\,\phi)\big \rangle_{1,x}\nonumber\\
&  \qquad=\,\frac{1}{2}\big \langle(\phi_{1})_{m}(-q^{-\zeta}t),\text{i}%
\partial_{A}\overset{x}{\triangleright}(\phi_{1}^{\ast})_{m}%
(t)\big \rangle_{1,x}\nonumber\\
&  \qquad\hspace{0.2in}+\frac{1}{2}\big \langle(\phi_{1}^{\ast})_{m}%
(t),\overline{\text{i}\partial_{A}}\,\overset{x}{\bar{\triangleright}}%
\,(\phi_{1})_{m}(-q^{-\zeta}t)\big \rangle_{1,x}\nonumber\\
&  \qquad=\,\int_{-\infty}^{+\infty}d_{1}\,p\,\frac{1}{2}\big (\,\overline
{(c_{1})_{p}(-q^{-\zeta}t)}\overset{p}{\circledast}p_{A}\overset
{p}{\circledast}(c_{1}^{\ast})_{\kappa^{-1}p}(\kappa^{2}t)\nonumber\\
&  \qquad\qquad\qquad\qquad\hspace{0.11in}+\,\overline{(c_{1}^{\ast}%
)_{\kappa^{-1}p}(\kappa^{2}t)}\overset{p}{\circledast}\overline{p_{A}}%
\overset{p}{\circledast}(c_{1})_{p}(-q^{-\zeta}t)\big )\nonumber\\
&  \qquad=\,\int_{-\infty}^{+\infty}d_{1}\,p\,\frac{1}{2}\big (\,\overline
{(c_{1})_{p}}\overset{p}{\circledast}p_{A}\overset{p}{\circledast}(c_{1}%
^{\ast})_{\kappa^{-1}p}\nonumber\\
&  \qquad\qquad\qquad\qquad\hspace{0.11in}+\,\overline{(c_{1}^{\ast}%
)_{\kappa^{-1}p}}\overset{p}{\circledast}\overline{p_{A}}\overset
{p}{\circledast}(c_{1})_{p}\big ), \label{ExpMom1}%
\end{align}
and%
\begin{align}
&  \big \langle\frac{1}{2}(P_{A}+\overline{P_{A}}\,)_{\phi}\big \rangle_{1,x}%
^{\prime}=\big \langle\frac{1}{2}(\phi\overset{x}{\triangleleft}(\text{i}%
\hat{\partial}_{A})+\phi\,\overset{x}{\bar{\triangleleft}}\,\overline
{\text{i}\hat{\partial}_{A}}\,),\phi\big \rangle_{1,x}^{\prime}\nonumber\\
&  \qquad=\,\frac{1}{2}\big \langle(\phi_{1})_{m}^{\prime}(q^{-\zeta
}t)\overset{x}{\triangleleft}(\text{i}\hat{\partial}_{A}),(\phi_{1}^{\ast
})_{m}^{\prime}(-t)\big \rangle_{1,x}^{\prime}\nonumber\\
&  \qquad\hspace{0.2in}+\frac{1}{2}\big \langle(\phi_{1}^{\ast})_{m}^{\prime
}(-t)\,\overset{x}{\bar{\triangleleft}}\,\overline{\text{i}\hat{\partial}_{A}%
}\,,(\phi_{1})_{m}^{\prime}(q^{-\zeta}t)\big \rangle_{1,x}^{\prime}\nonumber\\
&  \qquad=\,\int_{-\infty}^{+\infty}d_{1}\,p\,\frac{1}{2}\big ((c_{1}%
)_{p}^{\prime}(q^{-\zeta}t)\overset{p}{\circledast}p_{A}\overset
{p}{\circledast}\overline{(c_{1}^{\ast})_{\kappa^{-1}p}^{\prime}(-\kappa
^{2}t)}\nonumber\\
&  \qquad\qquad\qquad\qquad\hspace{0.11in}+\,(c_{1}^{\ast})_{\kappa^{-1}%
p}^{\prime}(-\kappa^{2}t)\overset{p}{\circledast}\overline{p_{A}}\overset
{p}{\circledast}\overline{(c_{1})_{p}^{\prime}(q^{-\zeta}t)}%
\,\big )\nonumber\\
&  \qquad=\,\int_{-\infty}^{+\infty}d_{1}\,p\,\frac{1}{2}\big ((c_{1}%
)_{p}^{\prime}\overset{p}{\circledast}p_{A}\overset{p}{\circledast}%
\overline{(c_{1}^{\ast})_{\kappa^{-1}p}^{\prime}}\nonumber\\
&  \qquad\qquad\qquad\qquad\hspace{0.11in}+\,(c_{1}^{\ast})_{\kappa^{-1}%
p}^{\prime}\overset{p}{\circledast}\overline{p_{A}}\overset{p}{\circledast
}\overline{(c_{1})_{p}^{\prime}}\,\big ). \label{ExpMom2}%
\end{align}
Furthermore, we have%
\begin{align}
&  \big \langle\frac{1}{2}(P_{A}+\overline{P_{A}}\,)_{\phi}\big \rangle_{2,x}%
=\big \langle\phi,\frac{1}{2}(\text{i}\hat{\partial}_{A}\,\overset{x}%
{\bar{\triangleright}}\,\phi+\overline{\text{i}\hat{\partial}_{A}}\overset
{x}{\triangleright}\phi)\big \rangle_{2,x}\nonumber\\
&  \qquad=\,\frac{1}{2}\big \langle(\phi_{2})_{m}(-q^{\zeta}t),\text{i}%
\hat{\partial}_{A}\,\overset{x}{\bar{\triangleright}}\,(\phi_{2}^{\ast}%
)_{m}(t)\big \rangle_{1,x}\nonumber\\
&  \qquad\hspace{0.2in}+\frac{1}{2}\big \langle(\phi_{2}^{\ast})_{m}%
(t),\overline{\text{i}\hat{\partial}_{A}}\overset{x}{\triangleright}(\phi
_{2})_{m}(-q^{\zeta}t)\big \rangle_{1,x}\nonumber\\
&  \qquad=\,\int_{-\infty}^{+\infty}d_{2}\,p\,\frac{1}{2}\big (\,\overline
{(c_{2})_{p}}\overset{p}{\circledast}p_{A}\overset{p}{\circledast}(c_{2}%
^{\ast})_{\kappa p}+\overline{(c_{2}^{\ast})_{\kappa p}}\overset
{p}{\circledast}\overline{p_{A}}\overset{p}{\circledast}(c_{2})_{p}%
\big ),\\[0.16in]
&  \big \langle\frac{1}{2}(P_{A}+\overline{P_{A}}\,)_{\phi}\big \rangle_{2,x}%
^{\prime}=\big \langle\frac{1}{2}(\phi\,\overset{x}{\bar{\triangleleft}%
}\,(\text{i}\partial_{A})+\phi\overset{x}{\triangleleft}\overline
{\text{i}\partial_{A}}\,),\phi\big \rangle_{2,x}^{\prime}\nonumber\\
&  \qquad=\,\frac{1}{2}\big \langle(\phi_{2})_{m}^{\prime}(q^{\zeta
}t)\,\overset{x}{\bar{\triangleleft}}\,(\text{i}\partial_{A}),(\phi_{2}^{\ast
})_{m}^{\prime}(-t)\big \rangle_{1,x}^{\prime}\nonumber\\
&  \qquad\hspace{0.2in}+\frac{1}{2}\big \langle(\phi_{2}^{\ast})_{m}^{\prime
}(-t)\overset{x}{\triangleleft}\overline{\text{i}\partial_{A}},(\phi_{2}%
)_{m}^{\prime}(q^{\zeta}t)\big \rangle_{1,x}^{\prime}\nonumber\\
&  \qquad=\,\int_{-\infty}^{+\infty}d_{2}\,p\,\frac{1}{2}\big ((c_{2}%
)_{p}^{\prime}\overset{p}{\circledast}p_{A}\overset{p}{\circledast}%
\overline{(c_{2}^{\ast})_{\kappa p}^{\prime}}+(c_{2}^{\ast})_{\kappa
p}^{\prime}\overset{p}{\circledast}\overline{p_{A}}\overset{p}{\circledast
}\overline{(c_{2})_{p}^{\prime}}\,\big ). \label{ExpMom4}%
\end{align}

The last expression in each of the above formulae shows us that expectation
values of momentum operators taken with respect to free-particle wave
functions are independent from time. To obtain this result we made use of the
fact that in (\ref{ExpMom1})-(\ref{ExpMom4}) the time-dependent phase factors
that are contained in the expansion coefficients [cf. the identities in
(\ref{CtCexp1}) and (\ref{CtCexp2})] commute with momentum variables and
cancel each other out.

For the sake of completeness we wish to write down expectation values for
position operators. We find%
\begin{align}
&  \big \langle\frac{1}{2}(X^{A}+\overline{X^{A}}\,)_{\phi}\big \rangle_{1,x}%
=\big \langle\phi,\frac{1}{2}(x^{A}+\overline{x^{A}}\,)\overset{x}%
{\circledast}\phi\big \rangle_{1,x}\nonumber\\
&  \qquad=\,\frac{1}{2}\big \langle(\phi_{1})_{m}(-q^{-\zeta}t),x^{A}%
\overset{x}{\circledast}(\phi_{1}^{\ast})_{m}(t)\big \rangle_{1,x}\nonumber\\
&  \qquad\hspace{0.2in}+\frac{1}{2}\big \langle(\phi_{1}^{\ast})_{m}%
(t),\overline{x^{A}}\overset{x}{\circledast}(\phi_{1})_{m}(-q^{-\zeta
}t)\big \rangle_{1,x}\nonumber\\
&  \qquad=\,\int_{-\infty}^{+\infty}d_{1}\,p\,\frac{1}{2}\big (\,\overline
{(c_{1})_{p}(-q^{-\zeta}t)}\overset{p}{\circledast}\big (\text{i}\partial
^{A}\,\overset{p}{\bar{\triangleright}}\,(c_{1}^{\ast})_{\kappa^{-1}p}%
(\kappa^{2}t)\big )\nonumber\\
&  \qquad\qquad\qquad\qquad\hspace{0.11in}+\,\overline{(c_{1}^{\ast}%
)_{\kappa^{-1}p}(\kappa^{2}t)}\overset{p}{\circledast}\big (\overline
{\,\text{i}\partial^{A}}\overset{p}{\triangleright}(c_{1})_{p}(-q^{-\zeta
}t)\big )\big ), \label{ExpPos1}%
\end{align}
and%
\begin{align}
&  \big \langle\frac{1}{2}(X^{A}+\overline{X^{A}}\,)_{\phi}\big \rangle_{1,x}%
^{\prime}=\big \langle\phi\overset{x}{\circledast}\frac{1}{2}(x^{A}%
+\overline{x^{A}}\,),\phi\big \rangle_{1,x}^{\prime}\nonumber\\
&  \qquad=\,\frac{1}{2}\big \langle(\phi_{1})_{m}^{\prime}(q^{-\zeta
}t)\overset{x}{\circledast}x^{A},(\phi_{1}^{\ast})_{m}^{\prime}%
(-t)\big \rangle_{1,x}^{\prime}\nonumber\\
&  \qquad\hspace{0.2in}+\frac{1}{2}\big \langle(\phi_{1}^{\ast})_{m}^{\prime
}(-t)\overset{x}{\circledast}\overline{x^{A}},(\phi_{1})_{m}^{\prime
}(q^{-\zeta}t)\big \rangle_{1,x}^{\prime}\nonumber\\
&  \qquad=\,\int_{-\infty}^{+\infty}d_{1}\,p\,\frac{1}{2}\big (\big ((c_{1}%
)_{p}^{\prime}(q^{-\zeta}t)\,\overset{p}{\bar{\triangleleft}}\,(\text{i}%
\partial^{A})\,\big )\overset{p}{\circledast}\overline{(c_{1}^{\ast}%
)_{\kappa^{-1}p}^{\prime}(-\kappa^{2}t)}\nonumber\\
&  \qquad\qquad\qquad\qquad\hspace{0.11in}+\,\big ((c_{1}^{\ast})_{\kappa
^{-1}p}^{\prime}(-\kappa^{2}t)\overset{p}{\triangleleft}\overline
{\text{i}\partial^{A}}\big )\overset{p}{\circledast}\overline{(c_{1}%
)_{p}^{\prime}(q^{-\zeta}t)}\,\big )^{\prime}. \label{ExpPos2}%
\end{align}
Likewise, we get%
\begin{align}
&  \big \langle\frac{1}{2}(X^{A}+\overline{X^{A}}\,)_{\phi}\big \rangle_{2,x}%
=\big \langle\phi,\frac{1}{2}(x^{A}+\overline{x^{A}}\,)\overset{x}%
{\circledast}\phi\big \rangle_{2,x}\nonumber\\
&  \qquad=\,\frac{1}{2}\big \langle(\phi_{2})_{m}(-t),x^{A}\overset
{x}{\circledast}(\phi_{2}^{\ast})_{m}(t)\big \rangle_{1,x}\nonumber\\
&  \qquad\hspace{0.2in}+\frac{1}{2}\big \langle(\phi_{2}^{\ast})_{m}%
(t),\overline{x^{A}}\overset{x}{\circledast}(\phi_{2})_{m}%
(-t)\big \rangle_{1,x}\nonumber\\
&  \qquad=\,\int_{-\infty}^{+\infty}d_{2}\,p\,\frac{1}{2}\big (\,\overline
{(c_{2})_{p}(-t)}\overset{p}{\circledast}\big (\text{i}\hat{\partial}%
^{A}\overset{p}{\triangleright}(c_{2}^{\ast})_{\kappa p}(t)\big )\nonumber\\
&  \qquad\qquad\qquad\qquad\hspace{0.11in}+\overline{(c_{2}^{\ast})_{\kappa
p}(t)}\overset{p}{\circledast}\big (\,\overline{\text{i}\hat{\partial}^{A}%
}\,\overset{p}{\bar{\triangleright}}\,(c_{2})_{p}(-t)\big )\big ),\\[0.16in]
&  \big \langle(X^{A}+\overline{X^{A}}\,)_{\phi}\big \rangle_{2,x}^{\prime
}=\big \langle\phi\overset{x}{\circledast}\frac{1}{2}(x^{A}+\overline{x^{A}%
}\,),\phi\big \rangle_{2,x}^{\prime}\nonumber\\
&  \qquad=\,\frac{1}{2}\big \langle(\phi_{2})_{m}^{\prime}(t)\overset
{x}{\circledast}x^{A},(\phi_{2}^{\ast})_{m}^{\prime}(-t)\big \rangle_{1,x}%
^{\prime}\nonumber\\
&  \qquad\hspace{0.2in}+\frac{1}{2}\big \langle(\phi_{2}^{\ast})_{m}^{\prime
}(-t)\overset{x}{\circledast}\overline{x^{A}},(\phi_{2})_{m}^{\prime
}(t)\big \rangle_{1,x}^{\prime}\nonumber\\
&  \qquad=\,\int_{-\infty}^{+\infty}d_{2}\,p\,\frac{1}{2}\big (\big ((c_{2}%
)_{p}^{\prime}(t)\overset{p}{\triangleleft}(\text{i}\hat{\partial}%
^{A})\big )\overset{p}{\circledast}\overline{(c_{2}^{\ast})_{\kappa p}%
^{\prime}(-t)}\nonumber\\
&  \qquad\qquad\qquad\qquad\hspace{0.11in}+\big ((c_{2}^{\ast})_{\kappa
p}^{\prime}(-t)\,\overset{p}{\bar{\triangleleft}}\,\overline{\text{i}%
\hat{\partial}^{A}}\,\big )\overset{p}{\circledast}\overline{(c_{2}%
)_{p}^{\prime}(t)}\,\big )^{\prime}. \label{ExpPos4}%
\end{align}

The arguments that showed us time-independence of expectation values of
momentum operators do not carry over to expectation values of position
operators. This should be rather clear, since in general a free particle does
not rest in space.

Once again, let us have a short look at the normalization conditions and
expectation values as they read for wave functions on momentum space [cf. Eqs.
(\ref{NorCon1})-(\ref{ExpPos4})]. From these expressions it should become
obvious that the probability densities for meeting a free particle at time $t$
in an eigenstate of the momentum operator are given by%
\begin{align}
(\rho_{1})_{\phi}(p_{A},t)  &  =\frac{1}{2}\big [\,\overline{(c_{1}%
)_{p}(-q^{-\zeta}t)}\overset{p}{\circledast}(c_{1}^{\ast})_{\kappa^{-1}%
p}(\kappa^{2}t)\nonumber\\
&  \hspace{0.15in}+\overline{(c_{1}^{\ast})_{\kappa^{-1}p}(\kappa^{2}%
t)}\overset{p}{\circledast}(c_{1})_{p}(-q^{-\zeta}t)\big ],\\[0.1in]
(\rho_{2})_{\phi}(p_{A},t)  &  =\frac{1}{2}\big [\,\overline{(c_{2}%
)_{p}(-q^{\zeta}t)}\overset{p}{\circledast}(c_{2}^{\ast})_{\kappa p}%
(\kappa^{-2}t)\nonumber\\
&  \hspace{0.15in}+\overline{(c_{2}^{\ast})_{\kappa p}(\kappa^{-2}t)}%
\overset{p}{\circledast}(c_{2})_{p}(-q^{\zeta}t)\big ],
\end{align}
or%
\begin{align}
(\rho_{1})_{\phi}^{\prime}(p_{A},t)  &  =\frac{1}{2}\big [(c_{1})_{p}^{\prime
}(q^{-\zeta}t)\overset{p}{\circledast}\overline{(c_{1}^{\ast})_{\kappa^{-1}%
p}^{\prime}(-\kappa^{2}t)}\nonumber\\
&  \hspace{0.15in}+(c_{1}^{\ast})_{\kappa^{-1}p}^{\prime}(-\kappa
^{2}t)\overset{p}{\circledast}\overline{(c_{1})_{p}^{\prime}(q^{-\zeta}%
t)}\,\big ],\\[0.1in]
(\rho_{2})_{\phi}^{\prime}(p_{A},t)  &  =\frac{1}{2}\big [(c_{2})_{p}^{\prime
}(q^{\zeta}t)\overset{p}{\circledast}\overline{(c_{2}^{\ast})_{\kappa
p}^{\prime}(-\kappa^{-2}t)}\nonumber\\
&  \hspace{0.15in}+(c_{2}^{\ast})_{\kappa p}(-\kappa^{-2}t)\overset
{p}{\circledast}\overline{(c_{2})_{p}^{\prime}(q^{\zeta}t)}\,\big ].
\end{align}
Of course, for a free particle there is no variation in\ the probability
densities with time, thus the time variable as argument can be dropped in the
above expressions.

Let us return to the expectation values of momentum operators. Their
independence from time tells us that momentum of a free particle is a constant
of motion. This becomes also evident from the Heisenberg equations of motion,
as they were introduced in part I:%
\begin{align}
\frac{d(P_{A})_{H}}{dt}  &  =\text{i}[H_{0},(P_{A})_{H}]=\text{i}%
P^{2}(2m)^{-1}\,(P_{A})_{H}-(P_{A})_{H}\,\text{i}P^{2}(2m)^{-1}=0,\nonumber\\
\frac{d(P_{A})_{H}^{\prime}}{dt}  &  =\text{i}[(P_{A})_{H}^{\prime}%
,H_{0}]=(P_{A})_{H}\,\text{i}(2m)^{-1}P^{2}-\text{i}(2m)^{-1}P^{2}%
\,(P_{A})_{H}=0, \label{HeiGleMom}%
\end{align}
where%
\begin{align}
(P_{A})_{H}  &  =\exp(\text{i}tH_{0})\,P_{A}\,\exp(-\text{i}tH_{0}%
)=P_{A},\nonumber\\
(P_{A})_{H}^{\prime}  &  =\exp(-\text{i}tH_{0})\,P_{A}\,\exp(\text{i}%
tH_{0})=P_{A}.
\end{align}
Notice that the commutators in (\ref{HeiGleMom}) vanish due to the property of
$H_{0}$ to be central in the algebra of momentum space.

It is rather instructive to write down expectation values of momentum and
position operators in the Heisenberg picture. To this end let us first
demonstrate how to obtain expectation values in the Heisenberg picture. In
complete analogy to the undeformed case we start from an expectation value in
the Schr\"{o}dinger picture and rewrite it in a way that wave functions become
independent from time:
\begin{align}
&  \big \langle\frac{1}{2}(P_{A}+\overline{P_{A}})_{\phi}\big \rangle_{1,x}%
=\nonumber\\
&  =\,\frac{1}{2}\big \langle(\phi_{1})_{m}(x^{B},-q^{-\zeta}t),P_{A}%
\overset{x}{\triangleright}(\phi_{1}^{\ast})_{m}(x^{C},t)\big \rangle_{1,x}%
\nonumber\\
&  \hspace{0.2in}+\frac{1}{2}\big \langle(\phi_{1}^{\ast})_{m}(x^{B}%
,t),\overline{P_{A}}\,\overset{x}{\bar{\triangleright}}\,(\phi_{1})_{m}%
(x^{C},q^{-\zeta}t)\big \rangle_{1,x}\nonumber\\
&  =\,\frac{1}{2}\big \langle(\phi_{1})_{m}(x^{B},0)\,\overset{x}%
{\bar{\triangleleft}}\,\exp(-\text{i}q^{-\zeta}tH_{0}),P_{A}\overset
{x}{\triangleright}\exp(-\text{i}tH_{0})\overset{x}{\triangleright}(\phi
_{1}^{\ast})_{m}(x^{C},0)\big \rangle_{1,x}\nonumber\\
&  \hspace{0.2in}+\frac{1}{2}\big \langle(P_{A}\exp(-\text{i}tH_{0}%
))\overset{x}{\triangleright}(\phi_{1}^{\ast})_{m}(x^{B},0),(\phi_{1}%
)_{m}(x^{C},0)\,\overset{x}{\bar{\triangleleft}}\,\exp(-\text{i}q^{-\zeta
}tH_{0})\big \rangle_{1,x}\nonumber\\
&  =\,\frac{1}{2}\big \langle\exp(-\text{i}tH_{0})\,\overset{x}{\bar
{\triangleright}}\,(\phi_{1})_{m}(x^{B},0),(P_{A}\exp(-\text{i}tH_{0}%
))\overset{x}{\triangleright}(\phi_{1}^{\ast})_{m}(x^{C},0)\big \rangle_{1,x}%
\nonumber\\
&  \hspace{0.2in}+\frac{1}{2}\big \langle(P_{A}\exp(-\text{i}tH_{0}%
))\overset{x}{\triangleright}(\phi_{1}^{\ast})_{m}(x^{B},0),\exp
(-\text{i}tH_{0})\,\overset{x}{\bar{\triangleright}}\,(\phi_{1})_{m}%
(x^{C},0)\big \rangle_{1,x}\nonumber\\
&  =\,\frac{1}{2}\big \langle(\phi_{1})_{m}(x^{B},0),(\exp(\text{i}%
tH_{0})P_{A}\exp(-\text{i}tH_{0}))\overset{x}{\triangleright}(\phi_{1}^{\ast
})_{m}(x^{C},0)\big \rangle_{1,x}\nonumber\\
&  \hspace{0.2in}+\frac{1}{2}\big \langle(\exp(\text{i}tH_{0})P_{A}%
\exp(-\text{i}tH_{0}))\overset{x}{\triangleright}(\phi_{1}^{\ast})_{m}%
(x^{B},0),(\phi_{1})_{m}(x^{C},0)\big \rangle_{1,x}\nonumber\\
&  =\,\frac{1}{2}\big \langle(\phi_{1})_{m}(x^{B},0),(P_{A})_{H}\overset
{x}{\triangleright}(\phi_{1}^{\ast})_{m}(x^{C},0)\big \rangle_{1,x}\nonumber\\
&  \hspace{0.2in}+\frac{1}{2}\big \langle(P_{A})_{H}\overset{x}{\triangleright
}(\phi_{1}^{\ast})_{m}(x^{B},0),(\phi_{1})_{m}(x^{C},0)\big \rangle_{1,x}%
\nonumber\\
&  =\,\frac{1}{2}\big \langle(\phi_{1})_{m}(x^{B},0),(P_{A})_{H}\overset
{x}{\triangleright}(\phi_{1}^{\ast})_{m}(x^{C},0)\big \rangle_{1,x}\nonumber\\
&  \hspace{0.2in}+\frac{1}{2}\big \langle(\phi_{1}^{\ast})_{m}(x^{B}%
,0),\overline{(P_{A})_{H}}\,\overset{x}{\bar{\triangleright}}\,(\phi_{1}%
)_{m}(x^{C},0)\big \rangle_{1,x}.
\end{align}

The first equality is the defining expression for\ the expectation value of a
momentum operator in the Schr\"{o}dinger picture. Then we introduce the
time-evolution operators by making use of the relations in (\ref{TimEntEbe1})
and (\ref{TimEntEbe4}). For the sake of convenience we rewrite the expression
in a way that all time evolution operators act from the left. Next, we use the
fact that the adjoints of the time-evolution operators are given by their
Hermitian conjugates. Finally, we are in a position to identify the
definitions of momentum operators in the Heisenberg picture.

Continuing these reasonings we find that in the Heisenberg picture expectation
values of momentum operators taken with respect to free-particle wave
functions are of the form%
\begin{align}
&  \big \langle\frac{1}{2}(P_{A}+\overline{P_{A}})_{\phi}\big \rangle_{1,x}%
=\nonumber\\
&  \qquad=\,\frac{1}{2}\big \langle(\phi_{1})_{m}(t=0),(P_{A})_{H}\overset
{x}{\triangleright}(\phi_{1}^{\ast})_{m}(t=0)\big \rangle_{1,x}\nonumber\\
&  \qquad\hspace{0.2in}+\frac{1}{2}\big \langle(\phi_{1}^{\ast})_{m}%
(t=0),\overline{(P_{A})_{H}}\,\overset{x}{\bar{\triangleright}}\,(\phi
_{1})_{m}(t=0)\big \rangle_{1,x}\nonumber\\
&  \qquad=\,\int_{-\infty}^{+\infty}d_{1}\,p\,\frac{1}{2}\big (\,\overline
{(c_{1})_{p}}\overset{p}{\circledast}p_{A}\overset{p}{\circledast}(c_{1}%
^{\ast})_{\kappa^{-1}p}\nonumber\\
&  \qquad\qquad\qquad\qquad\hspace{0.11in}+\,\overline{(c_{1}^{\ast}%
)_{\kappa^{-1}p}}\overset{p}{\circledast}\overline{p_{A}}\overset
{p}{\circledast}(c_{1})_{p}\big ),
\end{align}
and%
\begin{align}
&  \big \langle\frac{1}{2}(P_{A}+\overline{P_{A}})_{\phi}\big \rangle_{1,x}%
^{\prime}=\nonumber\\
&  \qquad=\,\frac{1}{2}\big \langle(\phi_{1})_{m}^{\prime}(t=0)\overset
{x}{\triangleleft}(P_{A})_{H}^{\prime},(\phi_{1}^{\ast})_{m}^{\prime
}(t=0)\big \rangle_{1,x}^{\prime}\nonumber\\
&  \qquad\hspace{0.2in}+\frac{1}{2}\big \langle(\phi_{1}^{\ast})_{m}^{\prime
}(t=0)\,\overset{x}{\bar{\triangleleft}}\,\overline{(P_{A})_{H}^{\prime}%
}\,,(\phi_{1})_{m}^{\prime}(t=0)\big \rangle_{1,x}^{\prime}\nonumber\\
&  \qquad=\,\int_{-\infty}^{+\infty}d_{1}\,p\,\frac{1}{2}\big ((c_{1}%
)_{p}^{\prime}\overset{p}{\circledast}p_{A}\overset{p}{\circledast}%
\overline{(c_{1}^{\ast})_{\kappa^{-1}p}^{\prime}}\nonumber\\
&  \qquad\qquad\qquad\qquad\hspace{0.11in}+\,(c_{1}^{\ast})_{\kappa^{-1}%
p}^{\prime}\overset{p}{\circledast}\overline{p_{A}}\overset{p}{\circledast
}\overline{(c_{1})_{p}^{\prime}}\,\big ).
\end{align}

Clearly, we can proceed in the same way for expectation values of position
operators. In the Heisenberg picture they read as%
\begin{align}
&  \big \langle\frac{1}{2}(X^{A}+\overline{X^{A}})_{\phi}\big \rangle_{1,x}%
=\nonumber\\
&  \qquad=\,\frac{1}{2}\big \langle(\phi_{1})_{m}(t=0),(X^{A})_{H}\overset
{x}{\circledast}(\phi_{1}^{\ast})_{m}(t=0)\big \rangle_{1,x}\nonumber\\
&  \qquad\hspace{0.2in}+\frac{1}{2}\big \langle(\phi_{1}^{\ast})_{m}%
(t=0),\overline{(X^{A})_{H}}\overset{x}{\circledast}(\phi_{1})_{m}%
(t=0)\big \rangle_{1,x}\nonumber\\
&  \qquad=\,\int_{-\infty}^{+\infty}d_{1}\,p\,\frac{1}{2}\big (\,\overline
{(c_{1})_{p}}\overset{p}{\circledast}\big ((X^{A})_{H}\,\overset{p}%
{\bar{\triangleright}}\,(c_{1}^{\ast})_{\kappa^{-1}p}\big )\nonumber\\
&  \qquad\qquad\qquad\qquad\hspace{0.11in}+\,\overline{(c_{1}^{\ast}%
)_{\kappa^{-1}p}}\overset{p}{\circledast}\big (\,\overline{(X^{A})_{H}%
}\overset{p}{\triangleright}(c_{1})_{p}\big )\big ),
\end{align}
and%
\begin{align}
&  \big \langle\frac{1}{2}(X^{A}+\overline{X^{A}})_{\phi}\big \rangle_{1,x}%
^{\prime}=\nonumber\\
&  \qquad=\,\frac{1}{2}\big \langle(\phi_{1})_{m}^{\prime}(t=0)\overset
{x}{\circledast}(X^{A})_{H}^{\prime},(\phi_{1}^{\ast})_{m}^{\prime
}(t=0)\big \rangle_{1,x}^{\prime}\nonumber\\
&  \qquad\hspace{0.2in}+\frac{1}{2}\big \langle(\phi_{1}^{\ast})_{m}^{\prime
}(t=0)\overset{x}{\circledast}\overline{(X^{A})_{H}^{\prime}},(\phi_{1}%
)_{m}^{\prime}(t=0)\big \rangle_{1,x}^{\prime}\nonumber\\
&  \qquad=\,\int_{-\infty}^{+\infty}d_{1}\,p\,\frac{1}{2}\big (\big ((c_{1}%
)_{p}^{\prime}\,\overset{p}{\bar{\triangleleft}}\,(X^{A})_{H}^{\prime
}\big )\overset{p}{\circledast}\overline{(c_{1}^{\ast})_{\kappa^{-1}p}%
^{\prime}}\nonumber\\
&  \qquad\qquad\qquad\qquad\hspace{0.11in}+\,\big ((c_{1}^{\ast})_{\kappa
^{-1}p}^{\prime}\overset{p}{\triangleleft}\overline{(X^{A})_{H}^{\prime}%
}\,\big )\overset{p}{\circledast}\overline{(c_{1})_{p}^{\prime}}%
\,\big )^{\prime},
\end{align}
where%
\begin{align}
(X_{A})_{H}  &  =\exp(\text{i}tH_{0})\,X_{A}\,\exp(-\text{i}tH_{0}%
),\nonumber\\
(X_{A})_{H}^{\prime}  &  =\exp(-\text{i}tH_{0})\,X_{A}\,\exp(\text{i}tH_{0}).
\end{align}
The expressions for the other geometries follow from the above formulae
through the substitutions in (\ref{SubGeo}).

We saw that expectation values of momentum operators are independent from
time, if they are taken with respect to free-particle wave functions. On the
contrary, expectation values of position operators should vary with time.
Again, this observation is in agreement with the Heisenberg equations of
motion for position operators,%
\begin{align}
\frac{d(X^{A})_{H}}{dt}  &  =\text{i}[H_{0},(X^{A})_{H}]=\text{i}H_{0}%
(X^{A})_{H}-\text{i}(X^{A})_{H}H_{0}\nonumber\\
&  =-(2m)^{-1}P^{2}\partial_{p}^{A}+\partial_{p}^{A}(2m)^{-1}P^{2}\nonumber\\
&  =-(\partial_{p}^{A})_{(2)}\big ((2m)^{-1}P^{2}\triangleleft(\partial
_{p}^{A})_{(1)}\big )+\partial_{p}^{A}(2m)^{-1}P^{2}\nonumber\\
&  =-\partial_{p}^{A}(2m)^{-1}P^{2}-(2m)^{-1}P^{2}\triangleleft\partial
_{p}^{A}+\partial_{p}^{A}(2m)^{-1}P^{2}\nonumber\\
&  =-(2m)^{-1}P^{2}\triangleleft\partial_{p}^{A}, \label{HeiGlImp1}%
\end{align}
and%
\begin{align}
\frac{d(X^{A})_{H}^{\prime}}{dt}  &  =\text{i}[(X^{A})_{H}^{\prime}%
,H_{0}]=\text{i}(X^{A})_{H}^{\prime}H_{0}-\text{i}H_{0}(X^{A})_{H}^{\prime
}\nonumber\\
&  =-\partial_{p}^{A}P^{2}(2m)^{-1}+P^{2}(2m)^{-1}\partial_{p}^{A}\nonumber\\
&  =-\big ((\partial_{p}^{A})_{(1)}\triangleright P^{2}(2m)^{-1}%
\big )(\partial_{p}^{A})_{(2)}+P^{2}(2m)^{-1}\partial_{p}^{A}\nonumber\\
&  =-\partial_{p}^{A}\triangleright P^{2}(2m)^{-1}-P^{2}(2m)^{-1}\partial
_{p}^{A}+P^{2}(2m)^{-1}\partial_{p}^{A}\nonumber\\
&  =-\partial_{p}^{A}\triangleright P^{2}(2m)^{-1}. \label{HeisGlImp2}%
\end{align}

For the fourth equality of both calculations we use the Leibniz rules of
partial derivatives on momentum space. Since they are determined by the
coproduct of partial derivatives, we write the Leibniz rules by using the
Sweedler notation for the coproduct. The fifth equality then is a consequence
of the trivial braiding of $H_{0}$.

To adjust the results in (\ref{HeiGlImp1}) and (\ref{HeisGlImp2}) to the
quantum spaces under consideration we need to know that

\begin{itemize}
\item[(i)] (braided line)%
\begin{align}
\partial^{1}\overset{p}{\triangleright}P^{2}(2m)^{-1}  &  =[[2]]_{q}%
P^{1}(2m)^{-1},\nonumber\\
\hat{\partial}^{1}\,\overset{p}{\bar{\triangleright}}\,P^{2}(2m)^{-1}  &
=[[2]]_{q^{-1}}P^{1}(2m)^{-1},\label{WirParHam1}\\[0.1in]
(2m)^{-1}P^{2}\overset{p}{\triangleleft}\hat{\partial}^{1}  &  =-[[2]]_{q^{-1}%
}(2m)^{-1}P^{1},\nonumber\\
(2m)^{-1}P^{2}\,\overset{p}{\bar{\triangleleft}}\,\partial^{1}  &
=-[[2]]_{q}(2m)^{-1}P^{1},
\end{align}

\item[(ii)] (q-deformed Euclidean space in three dimensions)%
\begin{align}
\partial^{A}\overset{p}{\triangleright}P^{2}(2m)^{-1}  &  =[[2]]_{q^{-2}}%
P^{A}(2m)^{-1},\nonumber\\
\hat{\partial}^{A}\,\overset{p}{\bar{\triangleright}}\,P^{2}(2m)^{-1}  &
=[[2]]_{q^{2}}P^{A}(2m)^{-1},\\[0.1in]
(2m)^{-1}P^{2}\overset{p}{\triangleleft}\hat{\partial}^{A}  &  =-[[2]]_{q^{2}%
}(2m)^{-1}P^{A},\nonumber\\
(2m)^{-1}P^{2}\,\overset{p}{\bar{\triangleleft}}\,\partial^{A}  &
=-[[2]]_{q^{-2}}(2m)^{-1}P^{A}. \label{WirParHam2}%
\end{align}

\end{itemize}

\noindent These relations can directly be derived from the Leibniz rules for
partial derivatives on braided line and q-deformed three-dimensional Euclidean
space (see part I of the paper). Notice that in the case of the braided line
the contravariant derivatives are identical with the covariant ones, while for
the q-deformed Euclidean space we have $\partial^{A}=g^{AB}\partial_{B}$.

Last but not least, we would like to mention that expressions with apostrophe
and those without apostrophe can be transformed into each other via
conjugation if we demand that
\begin{equation}
\overline{(c_{i})_{p}^{\prime}}=(c_{i})_{p},\quad\overline{(c_{i}^{\ast})_{p}%
}=(c_{i}^{\ast})_{p}^{\prime}.
\end{equation}
These identifications,\ in turn, imply that
\begin{equation}
\overline{(c_{i})_{p}^{\prime}(t)}=(c_{i})_{p}(t),\quad\overline{(c_{i}^{\ast
})_{p}(t)}=(c_{i}^{\ast})_{p}^{\prime}(t).
\end{equation}
Comparing the different expressions for probability densities and expectation
values [cf. Eqs. (\ref{NorCon1})-(\ref{ExpPos4})] should then tell us that
\begin{align}
\overline{(\rho_{i})_{\phi}(p_{A},t)}  &  =(\rho_{i})_{\phi}^{\prime}%
(p_{A},-t),\nonumber\\
\overline{\big \langle\frac{1}{2}(P_{A}+\overline{P_{A}})_{\phi}%
\big \rangle_{i,x}(t)}  &  =\big \langle\frac{1}{2}(P_{A}+\overline{P_{A}%
})_{\phi}\big \rangle_{i,x}^{\prime}(-t),\nonumber\\
\overline{\big \langle\frac{1}{2}(X^{A}+\overline{X^{A}})_{\phi}%
\big \rangle_{i,x}(t)}  &  =\big \langle\frac{1}{2}(X^{A}+\overline{X^{A}%
})_{\phi}\big \rangle_{i,x}^{\prime}(-t).
\end{align}
Finally, for the different expansions in terms of plane waves one can check
the identities%
\begin{equation}
\overline{(\phi_{i})_{m}(x^{A},t)}=(\phi_{i})_{m}^{\prime}(x^{A}%
,t),\quad\overline{(\phi_{i}^{\ast})_{m}(x^{A},t)}=(\phi_{i}^{\ast}%
)_{m}^{\prime}(x^{A},t). \label{ConWav}%
\end{equation}
If the reader is not familiar with conjugation properties of the objects of
q-analysis we recommend to consult Ref. \cite{qAn}.

\section{Theorem of Ehrenfest \label{TheEhr}}

In the last section we found solutions to q-deformed analogs of the
free-particle Schr\"{o}dinger equation. In what follows we would like to deal
with situations where the movement of a particle is influenced by the presence
of a rather weak potential, as it is the case in the theory of scattering.

Before we start developing a q-deformed version of propagator theory in part
III of the paper we would like to consider some more general aspects of a
system described by the Hamiltonian%
\begin{equation}
H=H_{0}+V(x^{A}).
\end{equation}
We demand that the potential $V(x^{A})$ is central in the algebra of
coordinate space and shows trivial braiding. These requirements ensure that
$H$ obeys the same algebraic properties as time derivatives. Furthermore, we
should have%
\begin{equation}
\overline{V(x^{A})}=V(x^{A})\quad\Rightarrow\quad\overline{H}=H.
\end{equation}

We wish to give examples for potentials with these features:

\begin{itemize}
\item[(i)] (braided line)%
\begin{equation}
V(x^{1})=-a|x^{1}|^{-b}, \label{PotBra}%
\end{equation}

\item[(ii)] (q-deformed Euclidean space in three dimensions)%
\begin{equation}
V(x^{A})=-ar^{-b}, \label{PotEuc}%
\end{equation}

\end{itemize}

\noindent where $r$ denotes the radius of q-deformed Euclidean space in three
dimensions. The constants $a\ $and $b$ have to be subject to

\begin{itemize}
\item[(i)] (braided line)%
\begin{align}
a\odot_{\bar{L}}f(p_{i})  &  =f(q^{-b}p_{1},p_{0})\otimes a,\nonumber\\
a\odot_{L}f(p_{i})  &  =f(q^{b}p_{1},p_{0})\otimes a,\\[0.1in]
a\odot_{\bar{L}}f(x^{i})  &  =f(q^{b}x^{1},x^{0})\otimes a,\nonumber\\
a\odot_{L}f(x^{i})  &  =f(q^{-b}x^{1},x^{0})\otimes a,
\end{align}

\item[(ii)] (q--deformed Euclidean space in three dimensions)%
\begin{align}
a\odot_{\bar{L}}f(p_{i})  &  =f(q^{-2b}p_{A},p_{0})\otimes a,\nonumber\\
a\odot_{L}f(p_{i})  &  =f(q^{2b}p_{A},p_{0})\otimes a,\\[0.1in]
a\odot_{\bar{L}}f(x^{i})  &  =f(q^{2b}x^{A},x^{0})\otimes a,\nonumber\\
a\odot_{L}f(x^{i})  &  =f(q^{-2b}x^{A},x^{0})\otimes a.
\end{align}

\end{itemize}

\noindent These relations guarantee for the trivial braiding of the potentials
in (\ref{PotBra}) and (\ref{PotEuc}). Additionally, we assume the element $a$
to be real and central in the algebra of position space.

To get a better understanding how the potential $V(x^{A})$ influences the
movement of a particle it is helpful to concentrate attention on the
Heisenberg equations for momentum operators, i.e.%
\begin{align}
\frac{d(P_{A})_{H}}{dt}  &  =\text{i}[H,(P_{A})_{H}]=\text{i}[H_{0}%
,(P_{A})_{H}]+\text{i}[V(x^{B}),(P_{A})_{H}]\nonumber\\
&  =\text{i}V(x^{B})\text{i}(\partial_{A})_{x}-\text{i}(\partial_{A}%
)_{x}\text{i}V(x^{A})\nonumber\\
&  =-V(x^{B})(\partial_{A})_{x}+([(\partial_{A})_{x}]_{(1)}\triangleright
V(x^{B}))[(\partial_{A})_{x}]_{(2)}\nonumber\\
&  =-V(x^{B})(\partial_{A})_{x}+(\partial_{A})_{x}\triangleright
V(x^{B})+V(x^{B})(\partial_{A})_{x}\nonumber\\
&  =(\partial_{A})_{x}\triangleright V(x^{B}),\label{HeiGleP1}\\[0.1in]
\frac{d(P_{A})_{H}^{\prime}}{dt}  &  =\text{i}[(P_{A})_{H}^{\prime
},H]=\text{i}[(P_{A})_{H}^{\prime},H_{0}]+\text{i}[(P_{A})_{H}^{\prime
},V(x^{B})]\nonumber\\
&  =\text{i}(\partial_{A})_{x}\text{i}V(x^{B})-\text{i}V(x^{B})\text{i}%
(\partial_{A})_{x}\nonumber\\
&  =-(\partial_{A})_{x}V(x^{B})+[(\partial_{A})_{x}]_{(2)}(V(x^{B}%
)\triangleleft\lbrack(\partial_{A})_{x}]_{(1)})\nonumber\\
&  =-(\partial_{A})_{x}V(x^{B})+(\partial_{A})_{x}V(x^{B})+V(x^{B}%
)\triangleleft(\partial_{A})_{x}\nonumber\\
&  =V(x^{B})\triangleleft(\partial_{A})_{x}. \label{HeiGleP2}%
\end{align}
For the third step of the above calculations we insert the operator
expressions for momentum in the Heisenberg picture. Notice that the
commutators with $H_{0}$ vanish due to the identities in (\ref{HeiGleMom}).
For the next step we apply the Leibniz rules for partial derivatives. Finally,
we make use of the trivial braiding of $V(x^{A}).$ This way, we arrive at
q-analogs of operator equations that correspond to the second law of Newtonian mechanics.

Next, we come to the Heisenberg equations for position operators. They take
the form%
\begin{align}
\frac{d(X^{A})_{H}}{dt}  &  =\text{i}[H,(X^{A})_{H}]=\text{i}[H_{0}%
,(X^{A})_{H}]+\text{i}[V(x^{B}),(X^{A})_{H}]\nonumber\\
&  =\partial_{p}^{A}\triangleright P^{2}(2m)^{-1},\label{HeiGleX1}\\[0.1in]
\frac{d(X^{A})_{H}^{\prime}}{dt}  &  =\text{i}[(X^{A})_{H}^{\prime
},H]=\text{i}[(X^{A})_{H}^{\prime},H_{0}]+\text{i}[(X^{A})_{H}^{\prime
},V(x^{B})]\nonumber\\
&  =(2m)^{-1}P^{2}\triangleleft\partial_{p}^{A}. \label{HeiGleX2}%
\end{align}
For these calculations we applied the results in (\ref{HeiGlImp1}) and
(\ref{HeisGlImp2}) from the proceeding section together with the property of
the potential $V(x^{A})$ to be central in the algebra of position space.
Notice that the above operator equations correspond to the definition of
momentum in classical mechanics.

Nothing prevents us from combining the Heisenberg equations for momentum
operators with those for position operators. In doing so, we can obtain

\begin{itemize}
\item[(i)] (braided line)%
\begin{align}
(\partial_{0})^{2}\overset{t}{\triangleright}(X^{1})_{H}  &  =\partial
_{0}\overset{t}{\triangleright}\big ((P^{2})_{H}(2m)^{-1}\,\overset{p}%
{\bar{\triangleleft}}\,\partial_{1}\big )\nonumber\\
&  =-[[2]]_{q}(\text{i}\partial_{1})\overset{x}{\triangleright}V(x^{B}%
)(2m)^{-1},\nonumber\\
(\hat{\partial}_{0})^{2}\,\overset{t}{\bar{\triangleright}}\,(X^{1})_{H}  &
=\hat{\partial}_{0}\,\overset{t}{\bar{\triangleright}}\,\big ((2m)^{-1}%
(P^{2})_{H}\overset{p}{\triangleleft}\hat{\partial}_{1}\big )\nonumber\\
&  =-[[2]]_{q^{-1}}(\text{i}\hat{\partial}_{1})\,\overset{x}{\bar
{\triangleright}}\,V(x^{B})(2m)^{-1},\label{OpeEqa1}\\[0.1in]
(X^{1})_{H}^{\prime}\overset{t}{\triangleleft}(\hat{\partial}_{0})^{2}  &
=-\big (\hat{\partial}_{1}\,\overset{p}{\bar{\triangleright}}\,(P^{2}%
)_{H}^{\prime}(2m)^{-1}\big )\overset{t}{\triangleleft}\hat{\partial}%
_{0}\nonumber\\
&  =[[2]]_{q^{-1}}(2m)^{-1}V(x^{B})\overset{x}{\triangleleft}(\text{i}%
\hat{\partial}_{1}),\nonumber\\
(X^{1})_{H}^{\prime}\,\overset{t}{\bar{\triangleleft}}\,(\partial_{0})^{2}  &
=-\big (\partial_{1}\overset{p}{\triangleright}(P^{2})_{H}^{\prime}%
(2m)^{-1}\big )\,\overset{t}{\bar{\triangleleft}}\,\partial_{0}\nonumber\\
&  =[[2]]_{q}(2m)^{-1}V(x^{B})\,\overset{x}{\bar{\triangleleft}}%
\,(\text{i}\partial_{1}),
\end{align}

\item[(ii)] (q-deformed Euclidean space in three dimensions)%
\begin{align}
(\partial_{0})^{2}\overset{t}{\triangleright}(X^{A})_{H}  &  =\partial
_{0}\overset{t}{\triangleright}\big ((P^{2})_{H}(2m)^{-1}\,\overset{p}%
{\bar{\triangleleft}}\,\partial^{A}\big )\nonumber\\
&  =-[[2]]_{q^{-2}}(\text{i}\partial^{A})\overset{x}{\triangleright}%
V(x^{B})(2m)^{-1},\nonumber\\
(\hat{\partial}_{0})^{2}\,\overset{t}{\bar{\triangleright}}\,(X^{A})_{H}  &
=\hat{\partial}_{0}\,\overset{t}{\bar{\triangleright}}\,\big ((2m)^{-1}%
(P^{2})_{H}\overset{p}{\triangleleft}\hat{\partial}^{A}\big )\nonumber\\
&  =-[[2]]_{q^{2}}(\text{i}\hat{\partial}^{A})\,\overset{x}{\bar
{\triangleright}}\,V(x^{B})(2m)^{-1},\\[0.1in]
(X^{A})_{H}^{\prime}\overset{t}{\triangleleft}(\hat{\partial}_{0})^{2}  &
=-\big (\hat{\partial}^{A}\,\overset{p}{\bar{\triangleright}}\,(P^{2}%
)_{H}^{\prime}(2m)^{-1}\big )\overset{t}{\triangleleft}\hat{\partial}%
_{0}\nonumber\\
&  =[[2]]_{q^{2}}(2m)^{-1}V(x^{B})\overset{x}{\triangleleft}(\text{i}%
\hat{\partial}^{A}),\nonumber\\
(X^{A})_{H}^{\prime}\,\overset{t}{\bar{\triangleleft}}\,(\partial_{0})^{2}  &
=-\big (\partial^{A}\overset{p}{\triangleright}(P^{2})_{H}^{\prime}%
(2m)^{-1}\big )\,\overset{t}{\bar{\triangleleft}}\,\partial_{0}\nonumber\\
&  =[[2]]_{q^{-2}}(2m)^{-1}V(x^{B})\,\overset{x}{\bar{\triangleleft}%
}\,(\text{i}\partial^{A}). \label{OpeEqa4}%
\end{align}

\end{itemize}

\noindent We first apply the results of (\ref{HeiGleX1}) and (\ref{HeiGleX2}).
Then we plug in the expressions listed in (\ref{WirParHam1})-(\ref{WirParHam2}%
). This way, we arrive at formulae that can be rewritten by making use of the
Heisenberg equations in (\ref{HeiGleP1}) and (\ref{HeiGleP2}).

Let us note that the above equations do not give the only possibility to
combine the Heisenberg equations of motions, but for us it seems to be the
most natural one. The reader should not be confused about the different time
derivatives. If we write $d/dt$ we mean the usual time derivative from
classical physics, which can be used to represent the partial derivatives
$\partial_{0}$ by%
\begin{gather}
\partial_{0}\overset{t}{\triangleright}f(x^{i})=\hat{\partial}_{0}%
\,\overset{t}{\bar{\triangleright}}\,f(x^{i})=df(x^{i})/dt,\nonumber\\
f(x^{i})\overset{t}{\triangleleft}\partial_{0}=f(x^{i})\,\overset{t}%
{\bar{\triangleleft}}\,\hat{\partial}_{0}=-df(x^{i})/dt.
\end{gather}

To get q-analogs of the celebrated Ehrenfest theorem it remains to take
expectation values of the operator equations in (\ref{OpeEqa1})-(\ref{OpeEqa4}%
). In this manner, we get%
\begin{align}
&  (\partial_{0})^{2}\overset{t}{\triangleright}\big \langle(\psi_{1}%
)_{m}(x^{B},0),(X^{A})_{H}\overset{x}{\triangleright}(\psi_{1}^{\ast}%
)_{m}(x^{C},0)\big \rangle_{1,x}=\nonumber\\
&  \quad=\,-\frac{[[2]]_{q^{-\zeta}}}{2m}\big \langle(\psi_{1})_{m}%
(x^{B},0),((\text{i}\partial^{A})\overset{x}{\triangleright}V(x^{D}%
))\overset{x}{\circledast}(\psi_{1}^{\ast})_{m}(x^{C},0)\big \rangle_{1,x}%
,\label{EhrThe1}\\[0.08in]
&  (\hat{\partial}_{0})^{2}\,\overset{t}{\bar{\triangleright}}%
\,\big \langle(\psi_{1}^{\ast})_{m}(x^{B},0),(X^{A})_{H}\,\overset{x}%
{\bar{\triangleright}}\,(\psi_{1})_{m}(x^{C},0)\big \rangle_{1,x}=\nonumber\\
&  \quad=\,-\frac{[[2]]_{q^{\zeta}}}{2m}\big \langle(\psi_{1}^{\ast}%
)_{m}(x^{B},0),((\text{i}\hat{\partial}^{A})\,\overset{x}{\bar{\triangleright
}}\,V(x^{D}))\overset{x}{\circledast}(\psi_{1})_{m}(x^{C},0)\big \rangle_{1,x}%
,
\end{align}
and%
\begin{align}
&  \big \langle(\psi_{1})_{m}^{\prime}(x^{B},0)\overset{x}{\triangleleft
}(X^{A})_{H}^{\prime},(\psi_{1}^{\ast})_{m}^{\prime}(x^{C}%
,0)\big \rangle_{1,x}^{\prime}\overset{t}{\triangleleft}(\hat{\partial}%
_{0})^{2}=\nonumber\\
&  \quad=\,\frac{[[2]]_{q^{\zeta}}}{2m}\big \langle(\psi_{1})_{m}^{\prime
}(x^{B},0)\overset{x}{\circledast}(V(x^{D})\overset{x}{\triangleleft}%
(\text{i}\hat{\partial}^{A})),(\psi_{1}^{\ast})_{m}^{\prime}(x^{C}%
,0)\big \rangle_{1,x}^{\prime},\\[0.08in]
&  \big \langle(\psi_{1}^{\ast})_{m}^{\prime}(x^{B},0)\,\overset{x}%
{\bar{\triangleleft}}\,(X^{A})_{H}^{\prime},(\psi_{1})_{m}^{\prime}%
(x^{C},0)\big \rangle_{1,x}^{\prime}\,\overset{t}{\bar{\triangleleft}%
}\,(\partial_{0})^{2}=\nonumber\\
&  \quad=\,\frac{[[2]]_{q^{-\zeta}}}{2m}\big \langle(\psi_{1}^{\ast}%
)_{m}^{\prime}(x^{B},0)\overset{x}{\circledast}(V(x^{D})\,\overset{x}%
{\bar{\triangleleft}}\,(\text{i}\partial^{A})),(\psi_{1})_{m}^{\prime}%
(x^{C},0)\big \rangle_{1,x}. \label{EhrThe4}%
\end{align}
Similar relations hold for the other geometries. As usual, they can be
obtained from the above formulae by applying the substitutions in
(\ref{SubGeo}). Thus, the details of their derivation are left to the reader.

Finally, it should be mentioned that the wave functions in respect to which
the expectation values in (\ref{EhrThe1})-(\ref{EhrThe4}) are taken have to be
subject to
\begin{align}
\text{i}\partial_{0}\overset{t}{\triangleright}(\psi_{1})_{m}^{\prime}(x^{i})
&  =H^{\prime}\overset{x}{\triangleright}(\psi_{1})_{m}^{\prime}%
(x^{i}),\nonumber\\
\text{i}\partial_{0}\overset{t}{\triangleright}(\psi_{1}^{\ast})_{m}(x^{i})
&  =H\overset{x}{\triangleright}(\psi_{1}^{\ast})_{m}(x^{i}%
),\label{SchrEqAlg1}\\[0.16in]
\text{i}\hat{\partial}_{0}\,\overset{t}{\bar{\triangleright}}\,(\psi_{2}%
)_{m}^{\prime}(x^{i})  &  =H^{\prime\prime}\,\overset{x}{\bar{\triangleright}%
}\,(\psi_{2})_{m}^{\prime}(x^{i}),\nonumber\\
\text{i}\hat{\partial}_{0}\,\overset{t}{\bar{\triangleright}}\,(\psi_{2}%
^{\ast})_{m}(x^{i})  &  =H\,\overset{x}{\bar{\triangleright}}\,(\psi_{2}%
^{\ast})_{m}(x^{i}),
\end{align}
and
\begin{align}
(\psi_{1})_{m}(x^{i})\,\overset{t}{\bar{\triangleleft}}\,(\text{i}\partial
_{0})  &  =(\psi_{1})_{m}(x^{i})\,\overset{x}{\bar{\triangleleft}}\,H^{\prime
},\nonumber\\
(\psi_{1}^{\ast})_{m}^{\prime}(x^{i})\,\overset{t}{\bar{\triangleleft}%
}\,(\text{i}\partial_{0})  &  =(\psi_{1}^{\ast})_{m}^{\prime}(x^{i}%
)\,\overset{x}{\bar{\triangleleft}}\,H,\\[0.1in]
(\psi_{2})_{m}(x^{i})\overset{t}{\triangleleft}(\text{i}\hat{\partial}_{0})
&  =(\psi_{2})_{m}(x^{i})\overset{x}{\triangleleft}H^{\prime\prime
},\nonumber\\
(\psi_{2}^{\ast})_{m}^{\prime}(x^{i})\overset{t}{\triangleleft}(\text{i}%
\hat{\partial}_{0})  &  =(\psi_{2}^{\ast})_{m}^{\prime}(x^{i})\overset
{x}{\triangleleft}H, \label{SchrEqAlg4}%
\end{align}
where
\begin{equation}
H^{\prime}=q^{-\zeta}H_{0}+V(x^{A}),\qquad H^{\prime\prime}=q^{\zeta}%
H_{0}+V(x^{A}).
\end{equation}

\section{Conservation of probability \label{ConProb}}

From classical quantum theory we know that the Schr\"{o}dinger equation
implies a continuity equation for the probability flux. If we assume that the
wave function behaves like a scalar and has trivial braiding we are able to
derive q-analogs of this continuity equation.

Towards this end we start from the probability densities $(i=1,2)$%
\begin{align}
(\rho_{i})_{m}(x^{A},t)=\,  &  \overline{(\psi_{i}^{\ast})_{m}(x^{A}%
,-t)}\overset{x,t}{\circledast}(\psi_{i})_{m}(x^{A},t),\nonumber\\
(\rho_{i}^{\ast})_{m}(x^{A},t)=\,  &  \overline{(\psi_{i})_{m}(x^{A}%
,-t)}\overset{x,t}{\circledast}(\psi_{i}^{\ast})_{m}(x^{A},t),
\label{DefProp1}\\[0.1in]
(\rho_{i})_{m}^{\prime}(x^{A},t)=\,  &  (\psi_{i})_{m}^{\prime}(x^{A}%
,t)\overset{x,t}{\circledast}\overline{(\psi_{i}^{\ast})_{m}^{\prime}%
(x^{A},-t)},\nonumber\\
(\rho_{i}^{\ast})_{m}^{\prime}(x^{A},t)=\,  &  (\psi_{i}^{\ast})_{m}^{\prime
}(x^{A},t)\overset{x,t}{\circledast}\overline{(\psi_{i})_{m}^{\prime}%
(x^{A},-t)}. \label{DefProp4}%
\end{align}
To ensure that all wave functions in these expressions describe the same
physical state we require for them to give solutions to the Schr\"{o}dinger
equations in (\ref{SchrEqAlg1})-(\ref{SchrEqAlg4}) in the sense that it holds
\begin{align}
(\psi_{1})_{m}^{\prime}(x^{i})  &  =\exp(-t\otimes\text{i}H^{\prime}%
)\overset{H^{\prime}|x}{\triangleright}\psi_{m}(x^{A},t=0),\nonumber\\
(\psi_{1}^{\ast})_{m}(x^{i})  &  =\exp(-t\otimes\text{i}H)\overset
{H|x}{\triangleright}\psi_{m}(x^{A},t=0),\\[0.1in]
(\psi_{2})_{m}^{\prime}(x^{i})  &  =\exp(-t\otimes\text{i}H^{\prime\prime
})\,\overset{H^{\prime\prime}|x}{\bar{\triangleright}}\,\psi_{m}%
(x^{A},t=0),\nonumber\\
(\psi_{2}^{\ast})_{m}(x^{i})  &  =\exp(-t\otimes\text{i}H)\,\overset{H|x}%
{\bar{\triangleright}}\,\psi_{m}(x^{A},t=0),
\end{align}
and%
\begin{align}
(\psi_{2})_{m}(x^{i})  &  =\psi_{m}(x^{A},t=0)\overset{x|H^{\prime\prime}%
}{\triangleleft}\exp(\text{i}H^{\prime\prime}\otimes t),\nonumber\\
(\psi_{2}^{\ast})_{m}^{\prime}(x^{i})  &  =\psi_{m}(x^{A},t=0)\overset
{x|H}{\triangleleft}\exp(\text{i}H\otimes t),\\[0.1in]
(\psi_{1})_{m}(x^{i})  &  =\psi_{m}(x^{A},t=0)\,\overset{x|H^{\prime}}%
{\bar{\triangleleft}}\,\exp(\text{i}H^{\prime}\otimes t),\nonumber\\
(\psi_{1}^{\ast})_{m}^{\prime}(x^{i})  &  =\psi_{m}(x^{A},t=0)\,\overset
{x|H}{\bar{\triangleleft}}\,\exp(\text{i}H\otimes t).
\end{align}

As next step we consider the time derivatives of the probability densities. In
doing so, we get, for example,
\begin{align}
&  \partial_{0}\overset{t}{\triangleright}(\rho_{1}^{\ast})_{m}(t)=\partial
_{0}\overset{t}{\triangleright}(\,\overline{(\psi_{1})_{m}(-t)}\overset
{x,t}{\circledast}(\psi_{1}^{\ast})_{m}(t))\nonumber\\
&  \quad=\,(\partial_{0}\overset{t}{\triangleright}\overline{(\psi_{1}%
)_{m}(-t)}\,)\overset{x,t}{\circledast}(\psi_{1}^{\ast})_{m}(t)+\overline
{(\psi_{1})_{m}(-t)}\overset{x,t}{\circledast}(\partial_{0}\overset
{t}{\triangleright}(\psi_{1}^{\ast})_{m}(t))\nonumber\\
&  \quad=\,\text{i}^{-1}(-H^{\prime}\overset{x}{\triangleright}\overline
{(\psi_{1})_{m}(-t)}\,)\overset{x,t}{\circledast}(\psi_{1}^{\ast}%
)_{m}(t)+\text{i}^{-1}\overline{(\psi_{1})_{m}(-t)}\overset{x,t}{\circledast
}(H\overset{x}{\triangleright}(\psi_{1}^{\ast})_{m}(t))\nonumber\\
&  \quad=\,-\text{i}^{-1}\big (q^{-\zeta}(2m)^{-1}P^{2}\overset{x}%
{\triangleright}\overline{(\psi_{1})_{m}(-t)}\,\big )\overset{x,t}%
{\circledast}(\psi_{1}^{\ast})_{m}(t)\nonumber\\
&  \quad\hspace{0.18in}\,\,+\text{i}^{-1}\overline{(\psi_{1})_{m}(-t)}%
\overset{x,t}{\circledast}\big ((2m)^{-1}P^{2}\overset{x}{\triangleright}%
(\psi_{1}^{\ast})_{m}(t)\big ). \label{DerConEq1}%
\end{align}
The first equality is the definition of the probability density and the second
equality is nothing other than the Leibniz rule of\ the time derivative. In
the third step we try to apply the Schr\"{o}dinger equations in
(\ref{SchrEqAlg1})-(\ref{SchrEqAlg4}). To achieve this we need the
Schr\"{o}dinger equation for the conjugate wave function $\overline{(\psi
_{1})_{m}(x^{A},-t)}$. It follows from the considerations%
\begin{align}
&  (\psi_{1})_{m}(x^{i})\,\overset{t}{\bar{\triangleleft}}\,(\text{i}%
\partial_{0})=(\psi_{1})_{m}(x^{i})\,\overset{x}{\bar{\triangleleft}%
}\,H^{\prime}\nonumber\\
\Rightarrow\quad &  -(\psi_{1})_{m}(x^{A},-t)\,\overset{t}{\bar{\triangleleft
}}\,(\text{i}\partial_{0})=(\psi_{1})_{m}(x^{A},-t)\,\overset{x}%
{\bar{\triangleleft}}\,H^{\prime}\nonumber\\
\Rightarrow\quad &  \overline{-(\psi_{1})_{m}(x^{A},-t)\,\overset{t}%
{\bar{\triangleleft}}\,(\text{i}\partial_{0})}=\overline{(\psi_{1})_{m}%
(x^{A},-t)\,\overset{x}{\bar{\triangleleft}}\,H^{\prime}}\nonumber\\
\Rightarrow\quad &  -(\text{i}\partial_{0})\overset{t}{\triangleright
}\,\overline{(\psi_{1})_{m}(x^{A},-t)}=H^{\prime}\,\overset{x}{\triangleright
}\,\overline{(\psi_{1})_{m}(x^{A},-t)}, \label{SchGleCon}%
\end{align}
where we made use of reality of the Hamiltonian $H^{\prime}$, i.e.%
\begin{equation}
\overline{H^{\prime}}=\overline{q^{-\zeta}P^{2}(2m)^{-1}}+\overline{V(x^{A}%
)}=q^{-\zeta}P^{2}(2m)^{-1}+V(x^{A})=H^{\prime},
\end{equation}
and the conjugation properties of the time derivative. For the last step in
(\ref{DerConEq1}) we insert the expressions for the Hamiltonians $H$ and
$H^{\prime}$. Realizing that the potential $V(x^{A})$ and the wave function
$(\psi_{1})_{m}(x^{A},-t)$ commute with each other (both behave like scalars)
we find that the contributions from $V(x^{A})$ cancel out against each other.

To proceed any further we need the identities
\begin{align}
&  \big (q^{-\zeta}(2m)^{-1}P^{2}\overset{x}{\triangleright}\overline
{(\psi_{1})_{m}(-t)}\,\big )\overset{x,t}{\circledast}(\psi_{1}^{\ast}%
)_{m}(t)=\nonumber\\
&  \quad=\,-\big (\,\overline{(\psi_{1})_{m}(-t)}\overset{x}{\triangleleft
}\partial^{A}\partial_{A}\big )\overset{x,t}{\circledast}(\psi_{1}^{\ast}%
)_{m}(t)(2m)^{-1}\nonumber\\
&  \quad=\,-\big [\big (\,\overline{(\psi_{1})_{m}(-t)}\overset{x}%
{\triangleleft}\partial^{A}\big )\overset{x,t}{\circledast}\big ((\partial
_{A})_{(1)}\triangleright(\psi_{1}^{\ast})_{m}(t)\big )\big ]\triangleleft
(\partial_{A})_{(2)}(2m)^{-1}\nonumber\\
&  \quad=\,-\big (\,\overline{(\psi_{1})_{m}(-t)}\overset{x}{\triangleleft
}\partial^{A}\big )\overset{x,t}{\circledast}\big (\partial_{A}\overset
{x}{\triangleright}(\psi_{1}^{\ast})_{m}(t)\big )(2m)^{-1}\nonumber\\
&  \quad\hspace{0.15in}\,\,-\big [\big (\,\overline{(\psi_{1})_{m}%
(-t)}\overset{x}{\triangleleft}\partial^{A}\big )\overset{x,t}{\circledast
}(\psi_{1}^{\ast})_{m}(t)\big ]\overset{x}{\triangleleft}\partial_{A}%
(2m)^{-1}, \label{Hil1}%
\end{align}
and%
\begin{align}
&  \overline{(\psi_{1})_{m}(-t)}\overset{x,t}{\circledast}\big ((2m)^{-1}%
P^{2}\overset{x}{\triangleright}(\psi_{1}^{\ast})_{m}(t)\big )=\nonumber\\
&  \quad=\,-\,\overline{(\psi_{1})_{m}(-t)}\overset{x,t}{\circledast
}\big (\partial^{A}\partial_{A}\overset{x}{\triangleright}(\psi_{1}^{\ast
})_{m}(t)\big )(2m)^{-1}\nonumber\\
&  \quad=\,-\,(\partial^{A})_{(2)}\triangleright\big [(\,\overline{(\psi
_{1})_{m}(-t)}\triangleleft(\partial^{A})_{(1)})\overset{x,t}{\circledast
}\big (\partial_{A}\overset{x}{\triangleright}(\psi_{1}^{\ast})_{m}%
(t)\big )\big ](2m)^{-1}\nonumber\\
&  \quad=\,\,-\,(\,\overline{(\psi_{1})_{m}(-t)}\overset{x}{\triangleleft
}\partial^{A})\overset{x,t}{\circledast}\big (\partial_{A}\overset
{x}{\triangleright}(\psi_{1}^{\ast})_{m}(t)\big )(2m)^{-1}\nonumber\\
&  \quad\hspace{0.16in}\,\,-\,\partial^{A}\overset{x}{\triangleright
}\big [\,\overline{(\psi_{1})_{m}(-t)}\overset{x,t}{\circledast}%
\big (\partial_{A}\overset{x}{\triangleright}(\psi_{1}^{\ast})_{m}%
(t)\big )\big ](2m)^{-1}. \label{Hil2}%
\end{align}
With the results of (\ref{Hil1}) and (\ref{Hil2}) the last expression in
relation (\ref{DerConEq1}) becomes%
\begin{align}
\partial_{0}\overset{t}{\triangleright}(\rho_{1}^{\ast})_{m}(t)=\,  &
\text{i}^{-1}\big [\big (\,\overline{(\psi_{1})_{m}(-t)}\overset
{x}{\triangleleft}\partial^{A}\big )\overset{x,t}{\circledast}(\psi_{1}^{\ast
})_{m}(t)\big ]\overset{x}{\triangleleft}\partial_{A}(2m)^{-1}\nonumber\\
\,  &  -\,\text{i}^{-1}\,\partial^{A}\overset{x}{\triangleright}%
\big [\,\overline{(\psi_{1})_{m}(-t)}\overset{x,t}{\circledast}\big (\partial
_{A}\overset{x}{\triangleright}(\psi_{1}^{\ast})_{m}(t)\big )\big ](2m)^{-1}.
\label{ZwiErg}%
\end{align}

Let us shortly explain the line of reasonings leading to (\ref{Hil1}) and
(\ref{Hil2}) In both calculations we first express momentum operators by
partial derivatives. In the calculation of (\ref{Hil1})\ we switch from the
left action of $H_{0}$ to its right action. In doing so, the factor
$q^{-\zeta}$ vanishes, while\ the wave functions remain unchanged, since they
behave as scalars with trivial braiding. One should also notice that we are
free to move the mass parameter $\ m$ to the far right. Then we apply Leibniz
rules for partial derivatives.\ These Leibniz rules can be simplified further
if we take into account once more that the wave functions transform as scalars.

Let us return to (\ref{ZwiErg}) and have a look on its right-hand side. It
remains to bring the partial derivative in the first summand from the far left
to the far right. This task can be achieved by means of the relation%
\begin{equation}
f\overset{x}{\triangleleft}\partial^{A}=S_{L}^{-1}(\partial^{A})\triangleright
f. \label{TraLinRecAct}%
\end{equation}
Remember that $S_{L}$ denotes the inverse of an antipode. It belongs to one of
the Hopf structures we could assign to the quantum spaces under consideration
[cf. part I of the article]. Exploiting relation (\ref{TraLinRecAct}) one can
verify that%
\begin{align}
\partial_{0}\overset{t}{\triangleright}(\rho_{1}^{\ast})_{m}(t)=\,  &
\text{i}^{-1}q^{-\zeta}\partial^{A}\overset{x}{\triangleright}%
\big [\big (\,\overline{(\psi_{1})_{m}(-t)}\overset{x}{\triangleleft}%
\partial_{A}\big )\overset{x,t}{\circledast}(\psi_{1}^{\ast})_{m}%
(t)\big ](2m)^{-1}\nonumber\\
\,  &  -\,\text{i}^{-1}\,\partial^{A}\overset{x}{\triangleright}%
\big [\,\overline{(\psi_{1})_{m}(-t)}\overset{x,t}{\circledast}\big (\partial
_{A}\overset{x}{\triangleright}(\psi_{1}^{\ast})_{m}(t)\big )\big ](2m)^{-1}.
\end{align}
From the last result we are able to read off the continuity equation%
\begin{equation}
\partial_{0}\overset{t}{\triangleright}(\rho_{1}^{\ast})_{m}(x^{B}%
,t)+\partial^{A}\overset{x}{\triangleright}[(j_{1}^{\ast})_{m}]_{A}%
(x^{B},t)=0,
\end{equation}
with%
\begin{align}
\lbrack(j_{1}^{\ast})_{m}]_{A}(x^{B},t)\equiv\,  &  \text{i}q^{-\zeta
}\big [\big (\,\overline{(\psi_{1})_{m}(x^{B},-t)}\overset{x}{\triangleleft
}\partial_{A}\big )\overset{x,t}{\circledast}(\psi_{1}^{\ast})_{m}%
(x^{B},t)\big ](2m)^{-1}\nonumber\\
\,  &  -\,\text{i}\,\big [\,\overline{(\psi_{1})_{m}(x^{B},-t)}\overset
{x,t}{\circledast}\big (\partial_{A}\overset{x}{\triangleright}(\psi_{1}%
^{\ast})_{m}(x^{B},t)\big )\big ](2m)^{-1}.
\end{align}

Repeating the same steps as above for the other q-geometries we finally get
the following collection of continuity equations:%
\begin{align}
&  \partial_{0}\overset{t}{\triangleright}(\rho_{1}^{\ast})_{m}(x^{B}%
,t)+\partial^{A}\overset{x}{\triangleright}[(j_{1}^{\ast})_{m}]_{A}%
(x^{B},t)=0,\nonumber\\
&  \hat{\partial}_{0}\,\overset{t}{\bar{\triangleright}}\,(\rho_{2}^{\ast
})_{m}(x^{B},t)+\hat{\partial}^{A}\,\overset{x}{\bar{\triangleright}}%
\,[(j_{2}^{\ast})_{m}]_{A}(x^{B},t)=0,\\[0.1in]
&  (\rho_{1})_{m}(x^{B},t)\,\overset{t}{\bar{\triangleleft}}\,\partial
_{0}+[(j_{1})_{m}]^{A}(x^{B},t)\,\overset{x}{\bar{\triangleleft}}%
\,\partial_{A}=0,\nonumber\\
&  (\rho_{2})_{m}(x^{B},t)\overset{t}{\triangleleft}\hat{\partial}_{0}%
+[(j_{2})_{m}]^{A}(x^{B},t)\overset{x}{\triangleleft}\hat{\partial}_{A}=0,
\end{align}
and%
\begin{align}
&  \partial_{0}\overset{t}{\triangleright}(\rho_{1}^{\ast})_{m}^{\prime}%
(x^{B},t)+\partial^{A}\overset{x}{\triangleright}[(j_{1}^{\ast})_{m}^{\prime
}]_{A}(x^{B},t)=0,\nonumber\\
&  \hat{\partial}_{0}\,\overset{t}{\bar{\triangleright}}\,(\rho_{2}^{\ast
})_{m}^{\prime}(x^{B},t)+\hat{\partial}^{A}\,\overset{x}{\bar{\triangleright}%
}\,[(j_{2}^{\ast})_{m}^{\prime}]_{A}(x^{B},t)=0,\\[0.1in]
&  (\rho_{1})_{m}^{\prime}(x^{B},t)\,\overset{t}{\bar{\triangleleft}%
}\,\partial_{0}+[(j_{1})_{m}^{\prime}]^{A}(x^{B},t)\,\overset{x}%
{\bar{\triangleleft}}\,\partial_{A}=0,\nonumber\\
&  (\rho_{2})_{m}^{\prime}(x^{B},t)\overset{t}{\triangleleft}\hat{\partial
}_{0}+[(j_{2})_{m}^{\prime}]^{A}(x^{B},t)\overset{x}{\triangleleft}%
\hat{\partial}_{A}=0.
\end{align}
The expressions for the probability fluxes take the form%
\begin{align}
\lbrack(j_{1}^{\ast})_{m}]_{A}(x^{B},t)\equiv\,  &  \frac{\text{i}q^{-\zeta}%
}{2m}\big [\big (\,\overline{(\psi_{1})_{m}(x^{B},-t)}\overset{x}%
{\triangleleft}\partial_{A}\big )\overset{x,t}{\circledast}(\psi_{1}^{\ast
})_{m}(x^{C},t)\big ]\nonumber\\
\,  &  -\,\frac{\text{i}}{2m}\big [\,\overline{(\psi_{1})_{m}(x^{B}%
,-t)}\overset{x,t}{\circledast}\big (\partial_{A}\overset{x}{\triangleright
}(\psi_{1}^{\ast})_{m}(x^{C},t)\big )\big ],\nonumber\\[0.06in]
\lbrack(j_{2}^{\ast})_{m}]_{A}(x^{B},t)\equiv\,  &  \frac{\text{i}q^{\zeta}%
}{2m}\,\big [\big (\,\overline{(\psi_{2})_{m}(x^{B},-t)}\,\overset{x}%
{\bar{\triangleleft}}\,\hat{\partial}_{A}\big )\overset{x,t}{\circledast}%
(\psi_{2}^{\ast})_{m}(x^{C},t)\big ]\nonumber\\
\,  &  -\,\frac{\text{i}}{2m}\big [\,\overline{(\psi_{2})_{m}(x^{B}%
,-t)}\overset{x,t}{\circledast}\big (\hat{\partial}_{A}\,\overset{x}%
{\bar{\triangleright}}\,(\psi_{2}^{\ast})_{m}(x^{C}%
,t)\big )\big ],\label{DefFlux1}\\[0.16in]
\lbrack(j_{1})_{m}]_{A}(x^{B},t)\equiv\,  &  \frac{\text{i}^{-1}q^{-\zeta}%
}{2m}\big [\,\overline{(\psi_{1}^{\ast})_{m}(x^{B},t)}\overset{x,t}%
{\circledast}\big (\partial_{A}\,\overset{x}{\bar{\triangleright}}\,(\psi
_{1})_{m}(x^{C},-t)\big )\big ]\nonumber\\
\,  &  -\,\frac{\text{i}^{-1}}{2m}\big [\big (\,\overline{(\psi_{1}^{\ast
})_{m}(x^{B},t)}\,\overset{x}{\bar{\triangleleft}}\,\partial_{A}%
\big )\overset{x,t}{\circledast}(\psi_{1})_{m}(x^{C}%
,-t)\big ],\nonumber\\[0.06in]
\lbrack(j_{2})_{m}]_{A}(x^{B},t)\equiv\,  &  -\,\frac{\text{i}^{-1}q^{\zeta}%
}{2m}\big [\,\overline{(\psi_{2}^{\ast})_{m}(x^{B},-t)}\overset{x,t}%
{\circledast}\big (\hat{\partial}_{A}\overset{x}{\triangleright}(\psi_{2}%
)_{m}(x^{C},t)\big )\big ]\nonumber\\
\,  &  +\,\frac{\text{i}^{-1}}{2m}\big [\big (\,\overline{(\psi_{2}^{\ast
})_{m}(x^{B},-t)}\overset{x}{\triangleleft}\hat{\partial}_{A}\big )\overset
{x,t}{\circledast}(\psi_{2})_{m}(x^{C},-t)\big ],
\end{align}
and, likewise,%
\begin{align}
\lbrack(j_{1}^{\ast})_{m}^{\prime}]_{A}(x^{B},t)\equiv\,  &  \frac
{\text{i}^{-1}q^{-\zeta}}{2m}\big [\big ((\psi_{1}^{\ast})_{m}^{\prime}%
(x^{B},-t)\overset{x}{\triangleleft}\partial_{A}\big )\overset{x,t}%
{\circledast}(\overline{\psi_{1})_{m}^{\prime}(x^{C},t)}\,\big ]\nonumber\\
\,  &  -\,\frac{\text{i}^{-1}}{2m}\big [(\psi_{1}^{\ast})_{m}^{\prime}%
(x^{B},-t)\overset{x,t}{\circledast}\big (\partial_{A}\overset{x}%
{\triangleright}\overline{(\psi_{1})_{m}^{\prime}(x^{C},t)}%
\,\big )\big ],\nonumber\\[0.06in]
\lbrack(j_{2}^{\ast})_{m}^{\prime}]_{A}(x^{B},t)\equiv\,  &  \frac
{\text{i}^{-1}q^{\zeta}}{2m}\big [\big ((\psi_{2}^{\ast})_{m}^{\prime}%
(x^{B},-t)\,\overset{x}{\bar{\triangleleft}}\,\hat{\partial}_{A}%
\big )\overset{x,t}{\circledast}\overline{(\psi_{2})_{m}^{\prime}(x^{C}%
,t)}\,\big ]\nonumber\\
\,  &  -\,\frac{\text{i}^{-1}}{2m}\big [(\psi_{2}^{\ast})_{m}^{\prime}%
(x^{B},-t)\overset{x,t}{\circledast}\big (\hat{\partial}_{A}\,\overset{x}%
{\bar{\triangleright}}\,\overline{(\psi_{2})_{m}^{\prime}(x^{C},t)}%
\,\big )\big ],\\[0.16in]
\lbrack(j_{1})_{m}^{\prime}]_{A}(x^{B},t)\equiv\,  &  \frac{\text{i}q^{-\zeta
}}{2m}\big [(\psi_{1})_{m}^{\prime}(x^{B},t)\overset{x,t}{\circledast
}\big (\partial_{A}\,\overset{x}{\bar{\triangleright}}\,\overline{(\psi
_{1}^{\ast})_{m}^{\prime}(x^{C},-t})\,\big )\big ]\nonumber\\
\,  &  -\,\frac{\text{i}}{2m}\big [\big ((\psi_{1})_{m}^{\prime}%
(x^{B},t)\,\overset{x}{\bar{\triangleleft}}\,\partial_{A}\big )\overset
{x,t}{\circledast}\overline{(\psi_{1}^{\ast})_{m}^{\prime}(x^{C}%
,-t)}\,\big ],\nonumber\\[0.06in]
\lbrack(j_{2})_{m}^{\prime}]_{A}(x^{B},t)\equiv\,  &  \,\frac{\text{i}%
q^{\zeta}}{2m}\big [(\psi_{2})_{m}^{\prime}(x^{B},t)\overset{x,t}{\circledast
}\big (\hat{\partial}_{A}\overset{x}{\triangleright}\overline{(\psi_{2}^{\ast
})_{m}^{\prime}(x^{C},-t)}\big )\big ]\nonumber\\
\,  &  -\,\frac{\text{i}}{2m}\big [\big ((\psi_{2})_{m}^{\prime}%
(x^{B},t)\overset{x}{\triangleleft}\hat{\partial}_{A}\big )\overset
{x,t}{\circledast}\overline{(\psi_{2}^{\ast})_{m}^{\prime}(x^{C},-t)}\,\big ].
\label{DefFlux4}%
\end{align}

Last but not least it should be noted that the different continuity equations
transform into each other via the operation of conjugation. This can be seen
if one takes into account that we have%
\begin{equation}
\overline{(\rho_{i})_{m}(x^{A},t)}=(\rho_{i})_{m}^{\prime}(x^{A}%
,t),\qquad\overline{(\rho_{i}^{\ast})_{m}(x^{A},t)}=(\rho_{i}^{\ast}%
)_{m}^{\prime}(x^{A},t),
\end{equation}
and%
\begin{equation}
\overline{\lbrack(j_{i})_{m}]_{A}(x^{B},t)}=[(j_{i})_{m}^{\prime}]^{A}%
(x^{B},t),\qquad\overline{[(j_{i}^{\ast})_{m}]_{A}(x^{B},t)}=[(j_{i}^{\ast
})_{m}^{^{\prime}}]^{A}(x^{B},t).
\end{equation}
These relations are a direct consequence of the identifications%
\[
\overline{(\psi_{i})_{m}(x^{A},t)}=(\psi_{i})_{m}^{\prime}(x^{A}%
,t),\qquad\overline{(\psi_{i}^{\ast})_{m}(x^{A},t)}=(\psi_{i}^{\ast}%
)_{m}^{\prime}(x^{A},t),
\]
and the defining expressions for probability density [cf. relations in
(\ref{DefProp1}) and (\ref{DefProp4})] and probability flux [cf. relations in
(\ref{DefFlux1})-(\ref{DefFlux4})].

\section{Conclusion\label{SecCon}}

Let us conclude our reasonings by some remarks. In part I of the paper we
worked out a mathematical and physical framework, which in part II was applied
to describe free-particles on q-deformed quantum spaces as the braided line
and the q-deformed Euclidean space in three dimensions. We introduced
q-analogs of the non-relativistic free-particle Hamiltonian and discussed
solutions to the corresponding Schr\"{o}dinger equations. We found that
q-deformed exponentials can be viewed as eigenfunctions of momentum and
energy. Furthermore, we saw that this set of functions is complete and
orthonormal. Then we extended the free-particle Hamiltonian by a potential and
showed that under certain assumptions we are able to formulate q-analogs of
the Ehrenfest theorem. Finally, we could prove continuity equations for
probability densities made up of wave functions with trivial braiding.

In this manner, our results seem to be in complete analogy to their undeformed
counterparts, to which they tend when the deformation parameter $q$ goes to 1.
However, there is one remarkable difference between a q-deformed theory and
its undeformed limit, since in a q-deformed theory we have to distinguish
different geometries. The reason for this lies in the fact that the braided
tensor category in which the expressions of our theory live is not uniquely
determined. This becomes more clear, if one realizes that each braided
category is characterized by a so-called braiding $\Psi$. The inverse
$\Psi^{-1}$ gives an equally good braiding, which leads to a second braided
category being different from the first one. This observation is reflected in
the occurrence of two\ differential calculi, different types of
q-exponentials, q-integrals and so on. In this manner each braided category
implies its own q-geometry, so we could write down different q-analogs of
well-known physical laws.

The point now is that we cannot restrict attention to one q-geometry, only,
since they are linked via the operation of conjugation \cite{OZ92, Maj94star,
Maj95star}. To be more precise, physical expressions have to be real, i.e.
invariant under the operation of conjugation. But this can only be achieved if
we combine expressions from different q-geometries. In the undeformed case,
however, the two categories become identical, so there is no necessity to take
account of different geometries.\vspace{0.16in}

\noindent\textbf{Acknowledgements}

First of all I am very grateful to Eberhard Zeidler for very interesting and
useful discussions, special interest in my work and financial support.
Furthermore, I would like to thank Alexander Schmidt for useful discussions
and his steady support. Finally, I thank Dieter L\"{u}st for kind hospitality.

\end{document}